\documentclass[aps,pra,reprint,superscriptaddress]{revtex4-1}
\usepackage{graphicx}
\usepackage{dcolumn}
\usepackage{amssymb}
\usepackage{amsmath}
\usepackage{siunitx}
\usepackage{color}

\DeclareSIUnit\Molar{\textsc{m}}

\makeatother

\begin{document}

\preprint{AIP/123-QED}

\title{Micro-rheology of a particle in a nonlinear bath: Stochastic Prandtl-Tomlinson model}

\author{Rohit Jain}
\email{rohit.jain@theorie.physik.uni-goettingen.de}
\affiliation{Institute for Theoretical Physics, Georg-August Universit\"{a}t G\"{o}ttingen, 37073 G\"{o}ttingen, Germany}

\author{F\'{e}lix Ginot}
\affiliation{Fachbereich Physik, Universit\"{a}t Konstanz, 78457 Konstanz, Germany}

\author{Matthias Kr\"{u}ger}
\affiliation{Institute for Theoretical Physics, Georg-August Universit\"{a}t G\"{o}ttingen, 37073 G\"{o}ttingen, Germany}

\date{\today}

\begin{abstract}
The motion of Brownian particles in nonlinear baths, such as, e.g., viscoelastic fluids, is of great interest. We theoretically study a simple model for such bath, where two particles are coupled via a sinusoidal potential. This model, which is an extension of the famous Prandtl Tomlinson model, has been found to reproduce some aspects of recent experiments, such as shear-thinning and position oscillations [J. Chem. Phys. {\bf 154}, 184904 (2021)]. Analyzing this model in detail, we show  that the predicted behavior of position oscillations agrees qualitatively with experimentally observed trends; (i) oscillations appear only in a certain regime of velocity and trap stiffness of the confining potential, and (ii), the amplitude and frequency of oscillations increase with driving velocity, the latter in a linear fashion. Increasing the potential barrier height of the model yields a rupture transition as a function of driving velocity, where the system abruptly changes from a mildly driven state to a strongly driven state. The frequency of oscillations scales as $(v_0-v_0^*)^{1/2}$ near the rupture velocity $v_0^*$, found for infinite trap stiffness. Investigating the (micro-)viscosity for different parameter ranges, we note that position oscillations leave their signature by an additional (mild) plateau in the flow curves, suggesting that oscillations influence the micro-viscosity. For a time-modulated driving, the mean friction force of the driven particle shows a pronounced resonance behavior, i.e, it changes strongly as a function of driving frequency. The model has two known limits: For infinite trap stiffness, it can be mapped to diffusion in a tilted periodic potential. For infinite bath friction, the original Prandtl Tomlinson model is recovered. We find that the flow curve of the model (roughly) crosses over between these two limiting cases.
\end{abstract}
\maketitle

 
\section{Introduction}
\label{sec:INTRO}
Optically trapped colloidal particles have been used to study rheological properties of complex fluids~\cite{yao2009microrheology, sriram2010active, Wilson2011-ip, Bechinger2014nonlinear, ahmed2015active, tassieri2015linear, Tassieri2016microrheology, robertson2018optical, Muller2020properties, madsen2021ultrafast, jain2021step}, to test the extension of fluctuation-dissipation relations in out-of-equilibrium systems such as in the context of stochastic thermodynamics~\cite{sekimoto1998langevin, Ciliberto2010Fluctuations, Bechinger2010FDT, seifert2012stochastic, Ciliberto2017Experiments} or to realize microscopic heat engines~\cite{Krishnamurthy2016HeatEngine, Bechinger2012HeatEngine}. In these applications, the optical trap can be used to apply well controlled and monitored forces on the probe. In the case of purely viscous fluids, such processes are described by Markovian Langevin dynamics~\cite{Lemons1997Langevin}. The Markovian limit, is, however, not applicable for more complex fluids, e.g. polymer or micellar solutions, colloidal suspensions or glass-forming liquids. Many nonlinear or non-Markovian properties of trapped particles in complex fluids have been observed and studied, experimentally and theoretically, with pronounced deviations from the behavior in purely viscous fluids ~\cite{gutsche2008colloids, Gazuz2009-er, Squires2005-mc, Bechinger2014nonlinear, Wilson2011-ip, Harrer2012-rs, Voigtmann2012force, Leitmann2013-wo, Benichou2013-yn,Winter2012-ja, Fuchs03, Berner2018Oscillating, Muller2020properties, jain2021step}. An example of particular interest is the observation of shear-thinning (decrease in viscosity with increasing shear rates) in viscoelastic fluids~\cite{Squires2005-mc, Fuchs2009active, Bechinger2014nonlinear, Voigtmann2012force}, which is also known from macro-rheological setups \cite{Larson1999structure}.
As a further consequence of the nonlinearity of viscoelastic fluids, it has been observed that the driven probe experiences an effective temperature that differs from the bath temperature~\cite{Wilson2011-ip, Demery2019driven} or it shows superdiffusive behavior~\cite{Benichou2013-yn, Winter2012-ja}. Some experiments have reported the occurrence of unsteady particle motion or oscillations when a colloid moves through a work-like micellar solution~\cite{Jayaraman2003Oscillations, Handzy2004Oscillatory, Berner2018Oscillating}. Theoretical studies have also predicted the dependence of viscosity on the trap stiffness, in case the bath particles are not subject to the trap potential \cite{Daldrop2017external, Kowalik2019memory, Muller2020properties, Muller2019thesis} or when the bath particles feel the trap directly~\cite{Lisy2019generalized, Tothova2021brownian}. The topic of nonlinear or nonequilibrium Langevin descriptions has been theoretically addressed in Refs.~\cite{zwanzig1973nonlinear, Nordholm1975systematic, Zwanzig2001-bd, Grabert2006projection, Kruger_2016, Meyer17, Vrugt2020projection}.

In recent experiments, oscillations of particle position have been observed for trapped colloidal particles driven in a worm-like  micellar solution~\cite{Berner2018Oscillating,jain2021step}. These oscillations are seen in the mean conditional displacements (MCD) when the colloidal particle is driven, in contrast to the equilibrium case, where MCD curves decay monotonically. This oscillatory dynamics can be reproduced using a generalized Langevin equation with friction memory term that is negative for long times \cite{Berner2018Oscillating}, which has been related to stress overshoots in macrorheology  \cite{Fuchs03} and, for microrheology, can induce persistent or ballistic  motion ~\cite{Zausch_2008,mitterwallner2020negative}. Less phenomenological, recent works introduced a simple \textit{bath-particle} model to investigate the experimental observations of shear thinning and oscillations~\cite{Muller2020properties,Muller2019thesis,jain2021step}. The model uses an interaction potential between the particle and bath, reminiscent of the celebrated Prandtl-Tomlinson (PT) model of dry friction~\cite{Prandtl1928, tomlinson1929}. Notably, this stochastic PT model describes shear-thinning and reproduces the mentioned oscillations. 

Here, we theoretically exploit the stochastic PT model~\cite{Muller2019thesis, Muller2020properties, jain2021step}, using Brownian dynamics simulations as well as analytical analysis in certain limits. We analyze the micro-viscosity, the amplitude and the frequency of oscillations as a function of driving velocity. A threshold velocity separates the near-equilibrium from the far from equilibrium regime:  the amplitude of oscillations starts to increase beyond this velocity,  shows a maximum, and then goes to zero for infinite driving velocity. The frequency of these oscillations is connected directly to the difference in average velocities of tracer and bath particles, and is found to scale approximately linearly with the driving speed $v_0$ (asymptotically exact for large $v_0$).
In the limit of large barrier height, a rupture transition is observed, where the above mentioned threshold velocity sharply separates states with zero from states with finite oscillation frequency.
The micro-viscosity, as a function of velocity, shows an additional (mild) plateau or  shoulder in the regime where oscillations are large. This coupling between viscosity and oscillations is investigated by application of a time modulated driving speed. Here, for states with pronounced oscillations under steady driving, the mean force shows a remarkable resonance behavior when driven modulated. 

The model can be analyzed analytically for infinite trap stiffness, including the mentioned rupture transition and the resonance behavior (the latter in a certain limit). In that case, the model maps to diffusion in a tilted potential. The model approaches the original Prandtl Tomlinson model for infinite bath friction. The observed flow curve is found to (roughly) cross over between these two limiting cases.

The manuscript is organised as follows: In Sec~\ref{PTModel}, we describe the stochastic PT model for a coupled tracer-bath system where the tracer particle is confined via an external harmonic potential. In Sec.~\ref{sec:SimulationsAndResults}, we perform Brownian dynamics simulations for this model system, where the trap moves at constant velocity. We discuss, in Sec.~\ref{sec:FrequencyAndOscillations}, the phenomenology of observed oscillations in our simulations and compare these to the existing experiments~\cite{Berner2018Oscillating}. In Sec~\ref{sec:Pin}, we discuss the limit of infinite trap stiffness, and, eventually, infinite potential barrier height. In Sec~\ref{sec:OscillationsIncreaseFriction}, we investigate the relation between oscillations and the mean friction force acting on the tracer particle, both for steady driving, as well as for time modulated driving. 

\section{Stochastic Prandtl-Tomlinson model}
\label{PTModel}
\begin{figure}[ht!]
	\centering
	\includegraphics[scale=0.25]{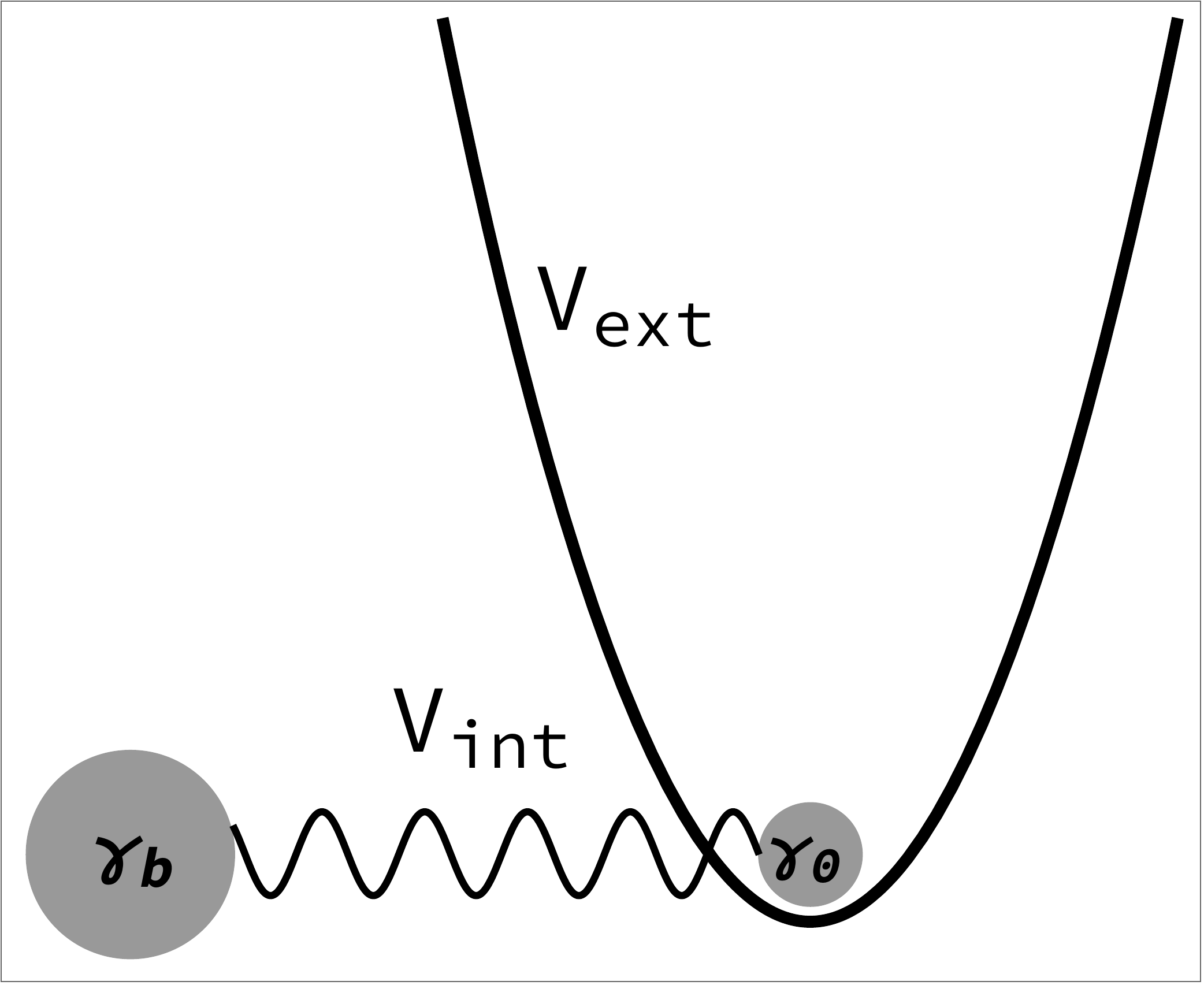}
	\caption{Schematic of the model system: A tracer particle is subject to an external potential, and connected to a bath particle. The external potential is then moved with velocity $v_0$, to drive the system out of equilibrium. }
	\label{SchematicPTModel}
\end{figure}

Consider a model of two coupled overdamped Brownian particles, interacting via a potential $V_{\textrm{int}}$ (see Fig.~\ref{SchematicPTModel}), in one spatial dimension. 
One of the particles plays the role of the tracer, which can be manipulated and observed in an experiment, the other plays the role of the bath, i.e., mimicking the degrees of freedom of the fluid which cannot be accessed (or which are not measured). This model gives rise to viscoelastic properties experienced by the tracer, and, in general introduces memory. For the case that  $V_{\textrm{int}}$ is harmonic, the resulting generalized Langevin equation for the tracer can be found analytically (see e.g.,~\cite{caldeira1981influence,Zwanzig2001-bd,Muller2020properties}). Here we choose a nonlinear coupling to allow for nonlinear effects, and states far from equilibrium. It has been argued earlier~\cite{Muller2020properties, jain2021step} that in order to observe shear thinning in this model, tracer and bath particles need to be unbound. Then they can move with different (mean) velocities, when driven, and the friction of the pair of two (corresponding to the viscosity) can vary. The details of this argument can be found in Ref.~\cite{Muller2020properties}. 
This leads us to the so-called stochastic Prandtl-Tomlinson (SPT) model, where $V_{\textrm{int}}$ is sinusoidal. Denoting $x$ and $q$ the positions of tracer and bath particle, respectively, we use
\begin{equation}
	V_{\textrm{int}} = -V_{0}\,\cos\left(\frac{2\pi}{d}(x-q)\right) \label{periodic_interaction_potential}.
\end{equation}
Here, $d$ introduces a length scale (a wavelength). The external potential is felt only by the tracer, and it is written as
\begin{equation}
V_{\textrm{ext}} = \frac{1}{2}\,\kappa(x-v_{0} t)^{2} , \label{external_potential}
\end{equation}
resembling a harmonic potential which is moving with velocity $v_{0}$, which can be time dependent. This corresponds to a realizable experimental protocol~\cite{Wilson2011-ip,Berner2018Oscillating,jain2021step}.

This leads to the following equations of motion for the two particles,
\begin{eqnarray}
\gamma_{0}\,\dot{x}(t) &=& - V'_{\textrm{ext}}(x(t)) -V'_{\textrm{int}}(x-q) + \eta_{0}(t), \label{eq:1}\\
\gamma_{b}\,\dot{q}(t) &=& V'_{int}(x-q) + \eta_{b}(t).  \label{eq:2}
\end{eqnarray}
$\gamma_{0}$ and $\gamma_{b}$ are the friction coefficients of tracer and bath particles, respectively, and ($\eta_{0}$, $\eta_{b}$) are the Gaussian, white, and independent random forces satisfying the fluctuation dissipation theorem, i.e.,
\begin{equation}
\left\langle\eta_{i}(t)\right\rangle = 0\,\text{and}\, \left\langle\eta_{i}(t)\eta_{j}(t')\right\rangle = 2k_{B}T\gamma_{i}\, \delta_{ij}\, \delta(t-t'). \label{noise}
\end{equation}
We denote this set of equations by stochastic  Prandtl Tomlinson model. The original Prandtl Tomlinson  model is one of the most popular models in the field of frictional physics, especially in nanotribology~\cite{PopovGray2014Prandtl,Prandtl1928,tomlinson1929,muser11a}. The overdamped version of it emerges from Eqs.~\eqref{eq:1} and \eqref{eq:2}, when setting $\gamma_b\to\infty$, so that the bath particle takes the role of a stationary background potential. A finite $\gamma_b$, as analyzed here, introduces (at least) one additional velocity scale, as we shall discuss below.

It is convenient to recast the equations of motion into a dimensionless form. We therefore rescale lengths in terms of $d$ in $V_{\textrm{int}}$ (see Eq.~(\ref{periodic_interaction_potential})). Energy is measured in units of $k_BT=\beta^{-1}$. Time will be measured in units of $\tau_{B}=\beta d^2 \gamma_{0}$, the time tracer particle takes to diffuse over the length scale $d$.  This yields the following unit-less variables and parameters,
\begin{align}
\bar{x} &= \frac{x}{d},\quad  \bar{t}=\frac{t}{\tau_{B}}, \quad
    \bar{\kappa} = \beta d^2\cdot\kappa, \quad \bar{\gamma}_{b} = \frac{\gamma_{b}}{\gamma_{0}}, \quad \bar{V}_{0} = \beta V_{0}, \notag \\ \bar{\eta}_{i} &= \beta d\cdot \eta_{i},  \quad \bar{v}_{0}=\beta\gamma_{0}d\cdot v_0. \label{scaling_parameters} 
\end{align}
In terms of these, the equations of motion become
\begin{eqnarray}
	\dot{\bar{x}} &=& -2\pi\bar{V}_{0} \sin\left(2\pi(\bar{x}-\bar{q})\right) - \bar{\kappa}\left(\bar{x} - \bar{v}_{0}\bar{t} \right) + \bar{\eta}_{0} \label{PT_eqn1},\\  
	\bar{\gamma}_{b}\,\dot{\bar{q}} &=& 2\pi\bar{V}_{0} \sin\left(2\pi(\bar{x}-\bar{q})\right) + \bar{\eta}_{b} \label{PT_eqn2},
\end{eqnarray}
with the random forces characterized by 
\begin{equation}\left\langle \bar{\eta}_{i}(\bar{t})\right\rangle = 0\,\,\,\text{and}\,\,\, \left\langle \bar{\eta}_{i}(\bar{t})\bar{\eta}_{j}(\bar{t}')\right\rangle = 2\bar{\gamma}_{i}\, \delta_{ij}\, \delta(\bar{t}-\bar{t}') .\label{noise-sorces}
\end{equation}
The system of coupled equations of motion described by~(\ref{PT_eqn1}) and~(\ref{PT_eqn2}) is the starting point of our computation. Solving the system of nonlinearly coupled stochastic equations analytically is rather challenging, and thus, we shall resort to Brownian dynamics simulations. However, for the special case of $\kappa\rightarrow\infty$, the steady state distribution can be evaluated exactly from these equations. We shall discuss this case in detail in Sec.~\ref{sec:Pin} and also partly in Sec.~\ref{sec:OscillationsIncreaseFriction}.
\section{Simulations: Flow curve, correlation function, and mean conditional displacement}
\label{sec:SimulationsAndResults}
\subsection{Method}
For the case of finite trap-stiffness, we deploy a stochastic Runge-Kutta method for weak convergence of order three to simulate the trajectories~\cite{Debrabant2010Runge-Kutta} using a time step $dt$ (see Table~\ref{table:simulation_parameters-set-1} in Appendix \ref{parameters} for simulation parameters). In order to remove any transient effects, we let the system reach the nonequilibrium steady state for a given velocity $v_0$, by waiting a time $t_{\textrm{eq}}$.  From the trajectories in steady state, the correlation function, mean-conditional displacement (MCD), and flow curve, are computed, and the sufficiency of the used values for $dt$ and $t_{\textrm{eq}}$ are checked. We shall refer to the same set of parameters for the rest of the paper unless specified otherwise.

\subsection{Flow curve}
In the steady state with driving, the tracer particle is displaced from the trap center due to friction forces of the bath. The mean friction force is shown in the so called flow curves, and found from  balancing with the restoring force of the harmonic trap. We  thus define the friction coefficient $\gamma$ as
\begin{equation}
  \gamma(v_{0}) =  \frac{\kappa|\left\langle x\right\rangle(t) - v_{0}t|}{v_{0}} . \label{def_flow-curve}
\end{equation}
$\left\langle x\right\rangle$ is the average particle position in the non-equilibrium steady state, which grows linearly in time. Note that $\gamma(v_{0})$ depends on drag-velocity due to the nonlinear nature of the potential $V_{\textrm{int}}$.
\begin{figure}
\centering
	\includegraphics[scale=0.4]{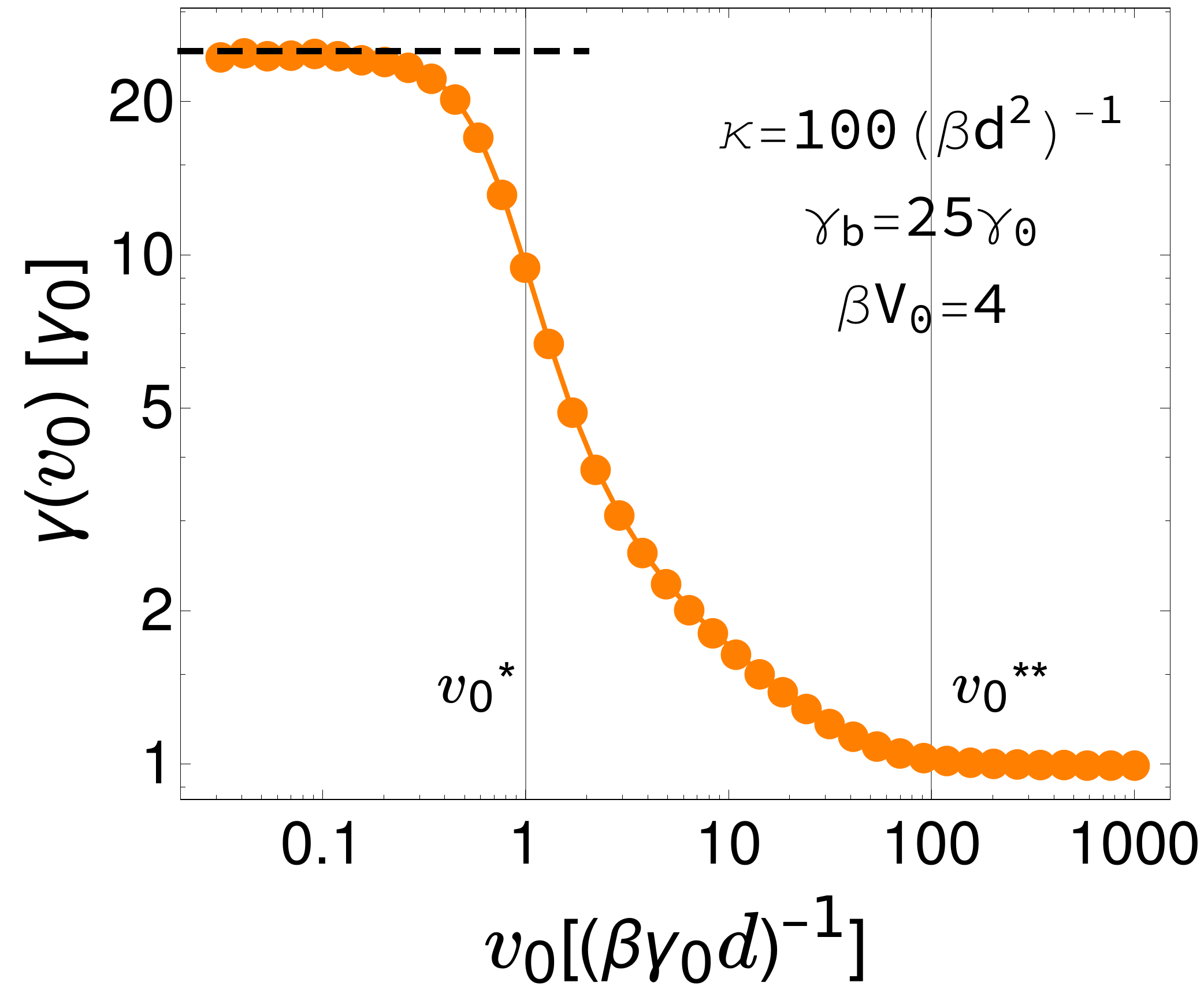}
	\caption{Flow curve $\gamma(v_0)$ of the stochastic PT model in units of the tracer’s friction coefficient $\gamma_0$. The parameters are given in the units discussed in Sec.~\ref{PTModel}. In the limit of small driving velocities, the flow curve asymptotically approaches the corresponding linear-response value (see Eq.~(\ref{LinearResponseResult}) in the text). For large driving velocities, the curve reduces to the tracer's friction coefficient $\gamma_0$.}
\label{fig:flowcurve}	
\end{figure}
In Fig.~\ref{fig:flowcurve}, we show the flow curve for the mentioned parameters as a function of drag-velocity $v_{0}$. For small driving velocities, $v_{0}\rightarrow 0$, the system shows a linear response regime, where the friction coefficient is independent of the shear rate. It can then be calculated from equilibrium fluctuations via the correlation function (see e.g. Ref.~\cite{Muller2020properties} for a derivation)
\begin{equation}
   \lim\limits_{v_{0}\rightarrow0} \gamma(v_{0}) = \beta\kappa^{2}\,\int_{0}^{\infty} dt\, \left\langle x(t)x(0)\right\rangle_{\textrm{eq}}.  \label{LinearResponseResult}
\end{equation}
The result of this relation is shown as the dashed line in Fig.~\ref{fig:flowcurve}. Naturally, for any $v_0$,  the friction is bound by the sum of bare friction coefficients, $\gamma(v_{0})\leq \gamma_0+\gamma_b$. Here, for small $v_0$, this bound is almost reached. Indeed, as will be analyzed in Sec.~\ref{sec:Pin}, the bound is reached in the limit of large values of potential barrier $V_0$ in Eq.~(\ref{periodic_interaction_potential}).

From the same potential barrier $V_0$, one can infer the maximal force between the tracer and the bath particle, i.e., $F_{max} = \frac{2\pi V_0} {d}$. If the force exceeds $F_{max}$, the bond between the tracer and bath particles breaks (a statement exact in absence of noise), and the velocity of the bath particle is on average smaller than the driving velocity $v_{0}$. Balancing the drag force with $F_{max}$, the critical velocity can be estimated, yielding $v_{0}^{\ast}\approx\frac{2\pi V_0} {\gamma_{b}d}$ (again, exact in absence of noise). Notably, this velocity scale does not exist in the original PT model (as $v_0^*\to0$ for $\gamma_b\to\infty$). Beyond the critical velocity $v_{0}^{\ast}$, the flow curve decreases, entering a shear-thinning regime. Eventually, in the limit of very large velocities, the flow curve reduces to the bare friction coefficient of the tracer, i.e. $\lim\limits_{v_{0}\rightarrow\infty}\gamma(v_{0}) = \gamma_{0}$. The flow curve presented here is typical of shear-thinning fluids  \cite{Squires2005-mc,Wilson2011-ip,jain2021step}, showing that this model may indeed capture important features of such systems.

A second velocity scale at $v_{0}^{**}=\kappa d/\gamma_0$,  which is also important in the original PT model, compares the relaxation time of the pure tracer, $\gamma_0/\kappa$ to the time $d/v_0$, which passes between two hoppings in the potential, if the bath particle is at rest.  In the units of the graph, it takes the value of $\bar v_0=100$, which is the scale at which the flow curve reaches its final value. We will further discuss this in Sec.~\ref{sec:OscillationsIncreaseFriction}.

\subsection{Correlation function and mean conditional displacement}
\label{subsec-CorrelationFunction_and_MCD}
In this subsection, we aim to study the particle’s fluctuations, inspired by the observation of oscillations in the mean-conditional displacement in references~\cite{Berner2018Oscillating, jain2021step}.
\begin{figure*}
        \begin{minipage}{0.3\textwidth}\textbf{\large{(a)}}
        \end{minipage}
    \hfill
        \begin{minipage}{0.3\textwidth}\textbf{\large{(b)}}
        \end{minipage}
    \hfill
        \begin{minipage}{0.3\textwidth}\textbf{\large{(c)}}
        \end{minipage}\\
        \includegraphics[width=0.32\textwidth]{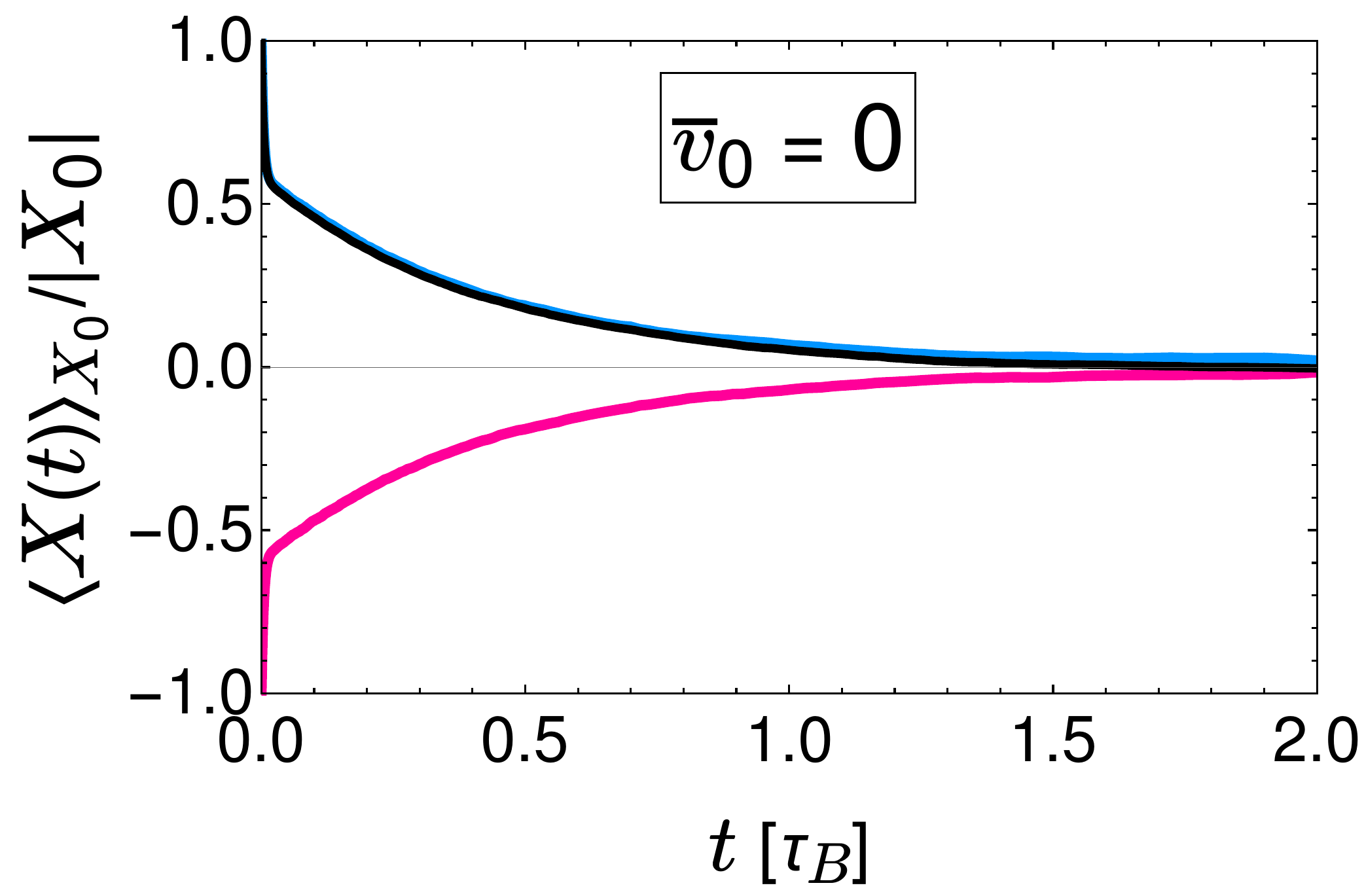}
    \hfill
	    \includegraphics[width=0.32\textwidth]{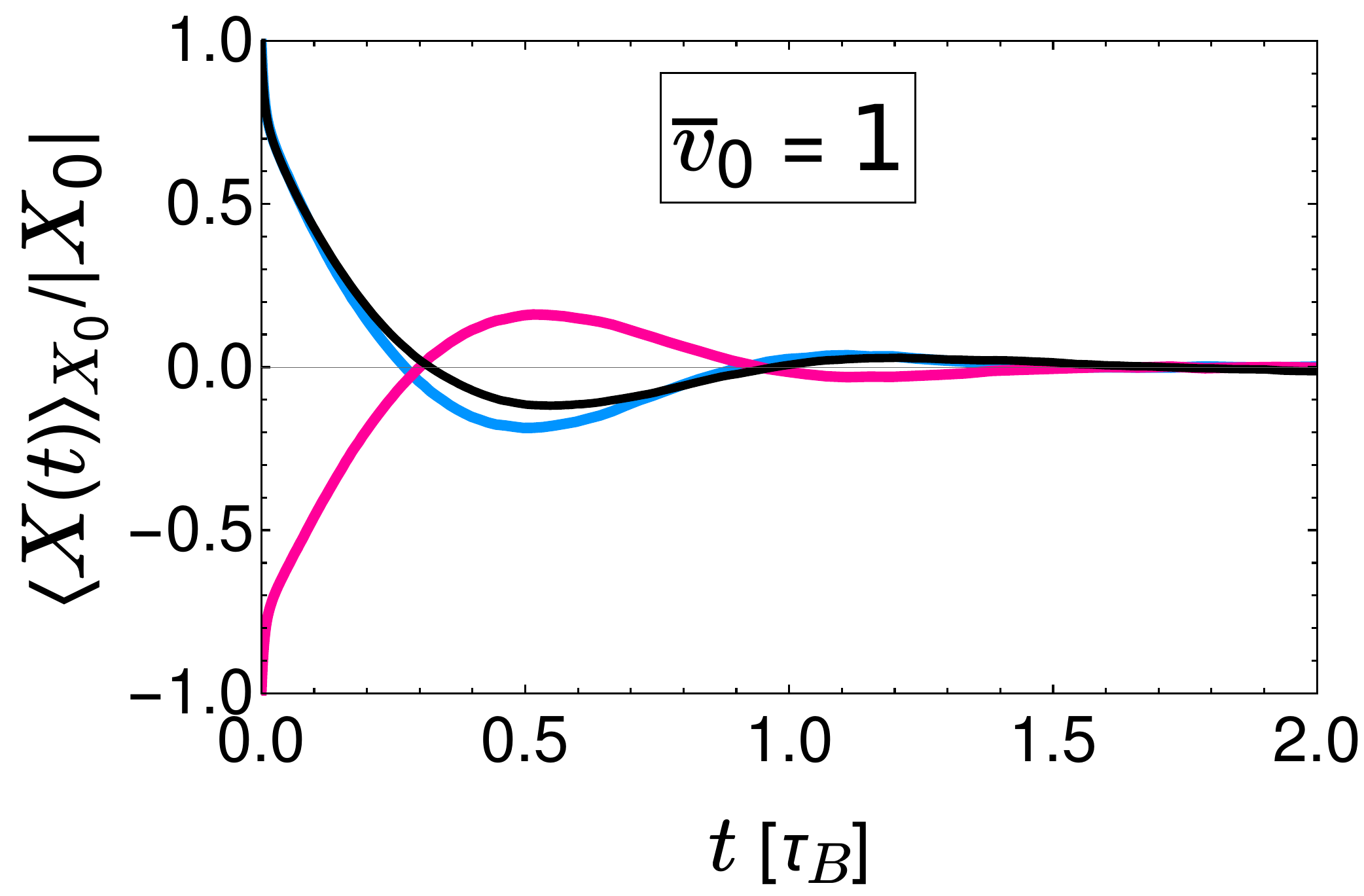}
    \hfill
     	\includegraphics[width=0.32\textwidth]{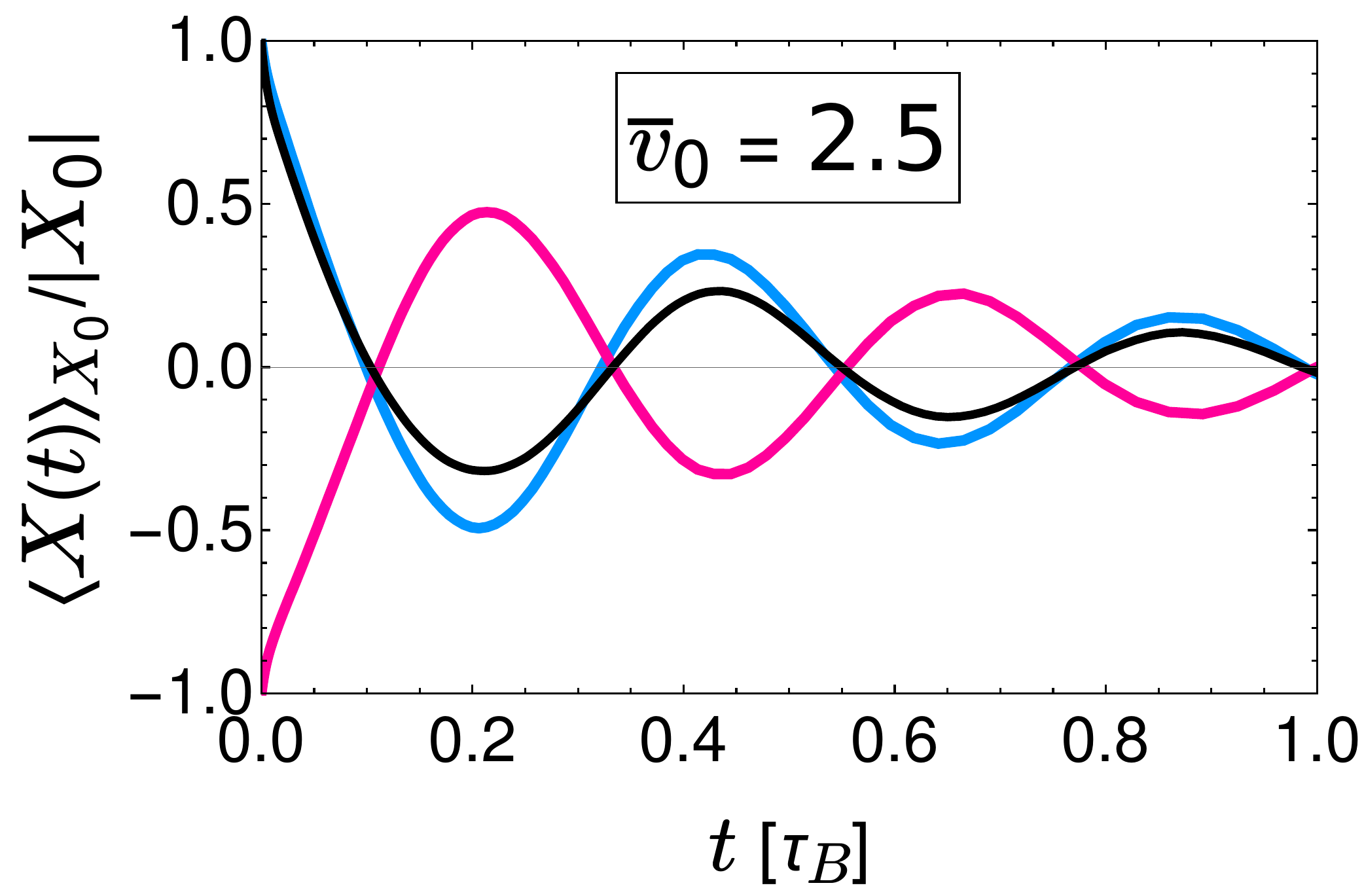}\\
        \begin{minipage}{0.3\textwidth}\textbf{\large{(d)}}
        \end{minipage}
    \hfill
        \begin{minipage}{0.3\textwidth}\textbf{\large{(e)}}
        \end{minipage}
    \hfill
        \begin{minipage}{0.3\textwidth}\textbf{\large{(f)}}
        \end{minipage}\\
	    \includegraphics[width=0.32\textwidth]{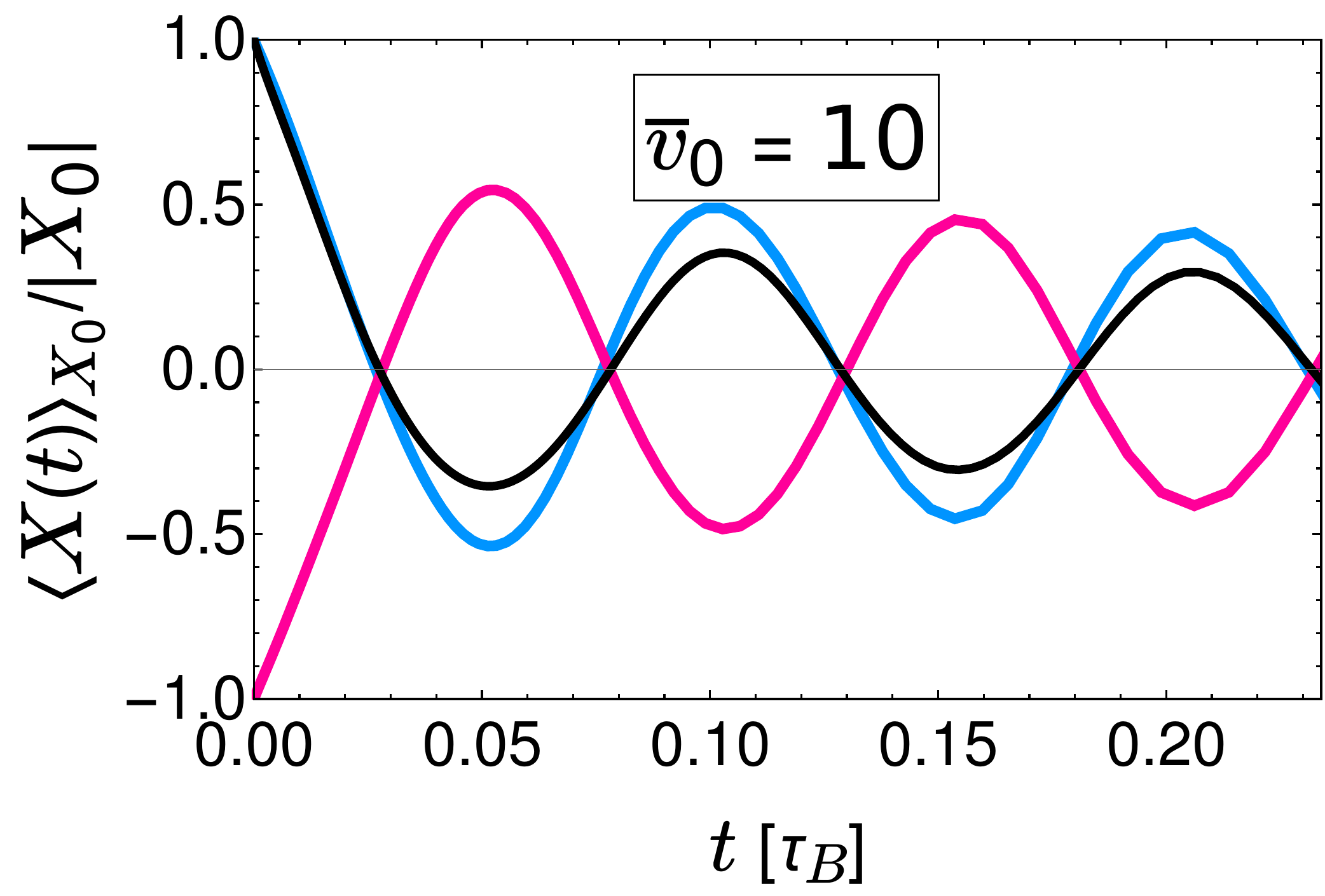}
	\hfill
	    \includegraphics[width=0.32\textwidth]{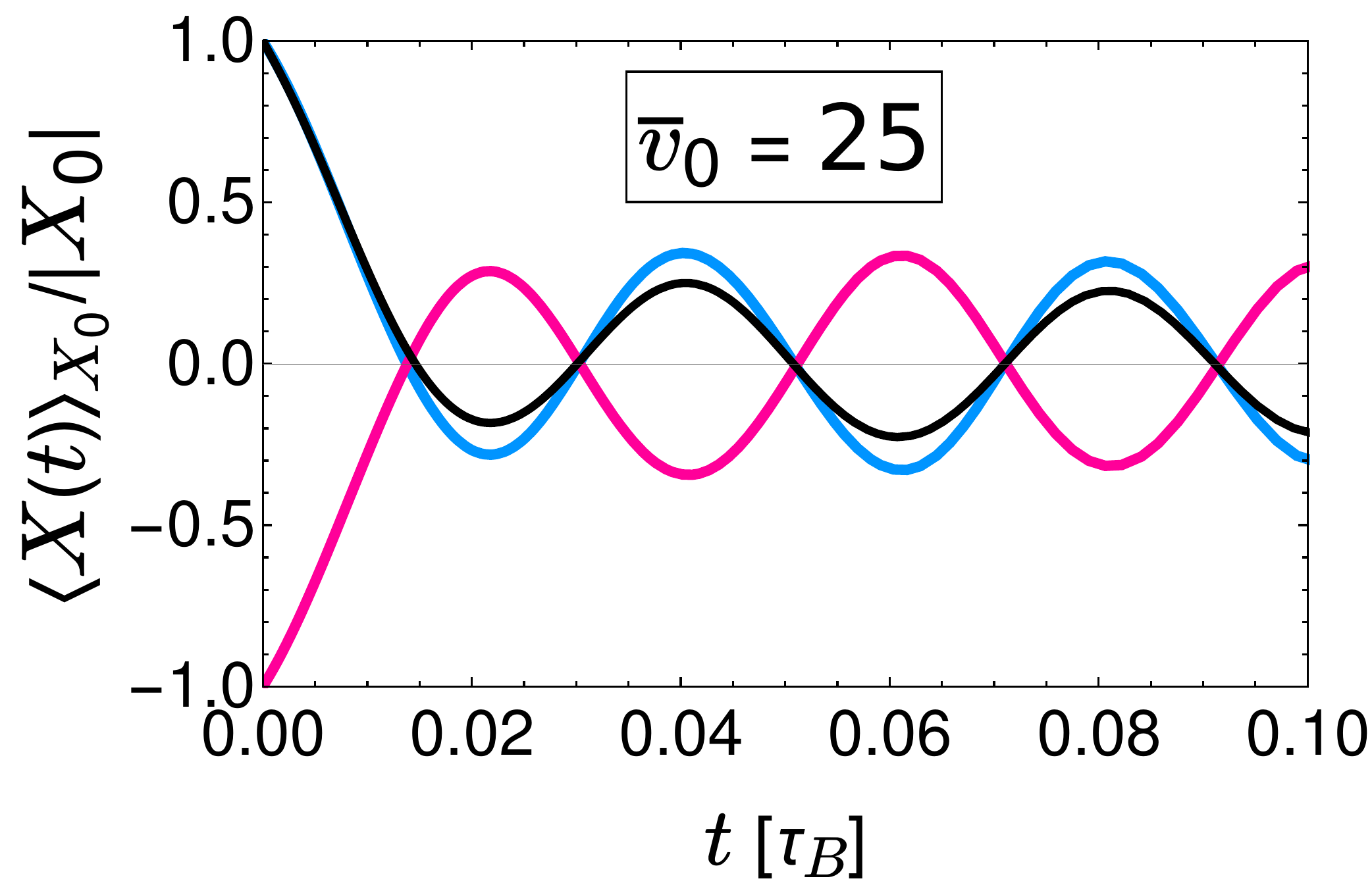}
	\hfill    
	    \includegraphics[width=0.32\textwidth]{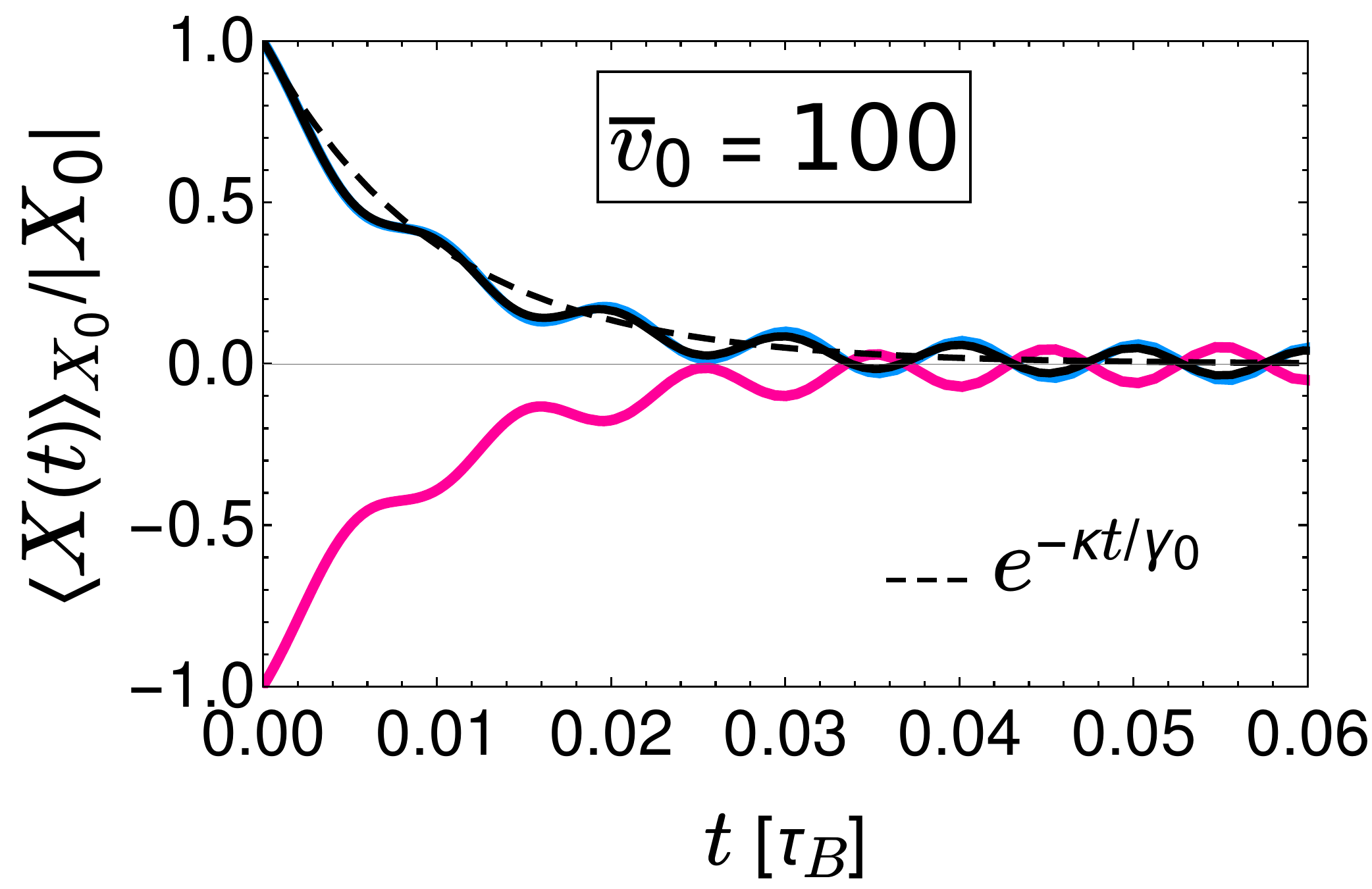}
	\caption{MCDs (light blue and light red for positive and negative values of $X_0$, respectively) and normalized correlation function (black) for different values of driving velocity with $\bar{V}_{0}=4$ and $\bar{\kappa}=100$. The dashed line in (f) shows $e^{-\kappa t/\gamma_0}$, i.e, the result of the system with $V_0=0$.}
    \label{fig-MCDs_and_CFs}
\end{figure*}
We define a new stochastic variable $X$ relative to the particle's mean position as $X= x-\left\langle x(t)\right\rangle$, such that the variable $X$ has zero mean, i.e. $\left\langle X(t)\right\rangle = 0$. Its fluctuations can then be quantified using the mean conditional displacement (MCD) defined as 
\begin{equation}
\left\langle X(t)\right\rangle_{X_0} = \int_{-\infty}^{\infty}dX\,X\, P(X,t|X_0,0), \label{def:MCD}
\end{equation}
where $P(X,t|X_0,0)$ is the conditional probability for the particle position  $X$ at time $t>0$, given the particle was at position $X_0$ at time $t=0$.

In Fig.~\ref{fig-MCDs_and_CFs}, we show the MCD curves in the stationary state as a function of time for different values of drag-velocity $v_{0}$. The parameters used here are the same as in the computation of flow curve of Fig.~~\ref{fig:flowcurve}, see Table~\ref{table:simulation_parameters-set-1} in Appendix~\ref{parameters}. In equilibrium, i.e. for $v_{0}=0$, the MCD decays monotonically to zero which is expected for any complex fluid made of overdamped particles~\cite{Berner2018Oscillating}, as the corresponding Fokker Planck operator has purely real eigenvalues. However, a qualitatively different behavior is observed for the case when the particle is driven with finite velocities. In the nonequilibrium steady-state, the MCDs do not decay monotonically, but rather show oscillations, with frequencies increasing with increasing driving velocity. We shall study the dependence of amplitude and frequency of the oscillations on $v_{0}$ in more details in the next section. 

In equilibrium, MCD and the correlation function are related by
\begin{align}
    \frac{\left\langle X(t)\right\rangle_{X_0}}{X_0}=\frac{\left\langle X(t)X(0)\right\rangle}{\langle X^2\rangle},
\label{eq:MCDC}
\end{align}
stating that the correlation function and MCD differ only by the overall amplitude. Eq.~\eqref{eq:MCDC} holds if $X$ is described by a linear Langevin equation, which is true in equilibrium~\cite{zwanzig1961lectures, mori1965transport}, but may not be so for the driven states, because the system of equations~(\ref{PT_eqn1}) and~(\ref{PT_eqn2}) is nonlinear. To test this statement beyond equilibrium, we  compare in Fig.~\ref{fig-MCDs_and_CFs} the MCD (blue and red lines) and the correlation function (black line) for the driven states as well. For the range of velocities used in our simulations, the two quantities partly deviate, but follow each other qualitatively. Indeed, the MCD for positive and negative values of $X_0$ are for the driven cases not perfectly symmetric, which is another hint that the underlying description for $X$ is nonlinear.

Due to the given similarity, we shall in the following use the correlation function rather than MCD to analyze amplitude and frequency of oscillations, as it is easier to extract from the simulated trajectories (MCD requires division by $X_0$, which can be small). In case of visible nonlinear effects, the function $\left\langle X(t)\right\rangle_{X_0}/{X_0}$ may strictly not be independent of $X_0$, so that the correlation function is also a better defined quantity.

The origin of the observed oscillations is rooted in the fact that in the (nonequilibrium) steady-state, for velocities larger than the critical velocity mentioned above, the colloid and the bath particles travel with different velocities~\cite{jain2021step}, i.e., the bath  particle then moves with an average speed  smaller than $v_{0}$. The tracer  is thus subject to a periodic force, which results in the seen oscillations. In Sec.~\ref{subsec-AmplitudeAndFrequency}, we demonstrate how the difference in velocities can be exploited to calculate the frequency of oscillations.

While the phenomenon of oscillations appears quite natural in this model, the agreement of the model with experimental data is somewhat surprising. The next section is devoted to comparing the model's predictions to experiments \cite{Berner2018Oscillating} on a qualitative level. 

\section{Phenomenology of Oscillations and comparison to Experiments~\cite{Berner2018Oscillating}}
\label{sec:FrequencyAndOscillations}

\subsection{Amplitude of oscillations} 
\label{subsec-AmplitudeAndFrequency}
\begin{figure}
    \centering
    \includegraphics[scale=0.3]{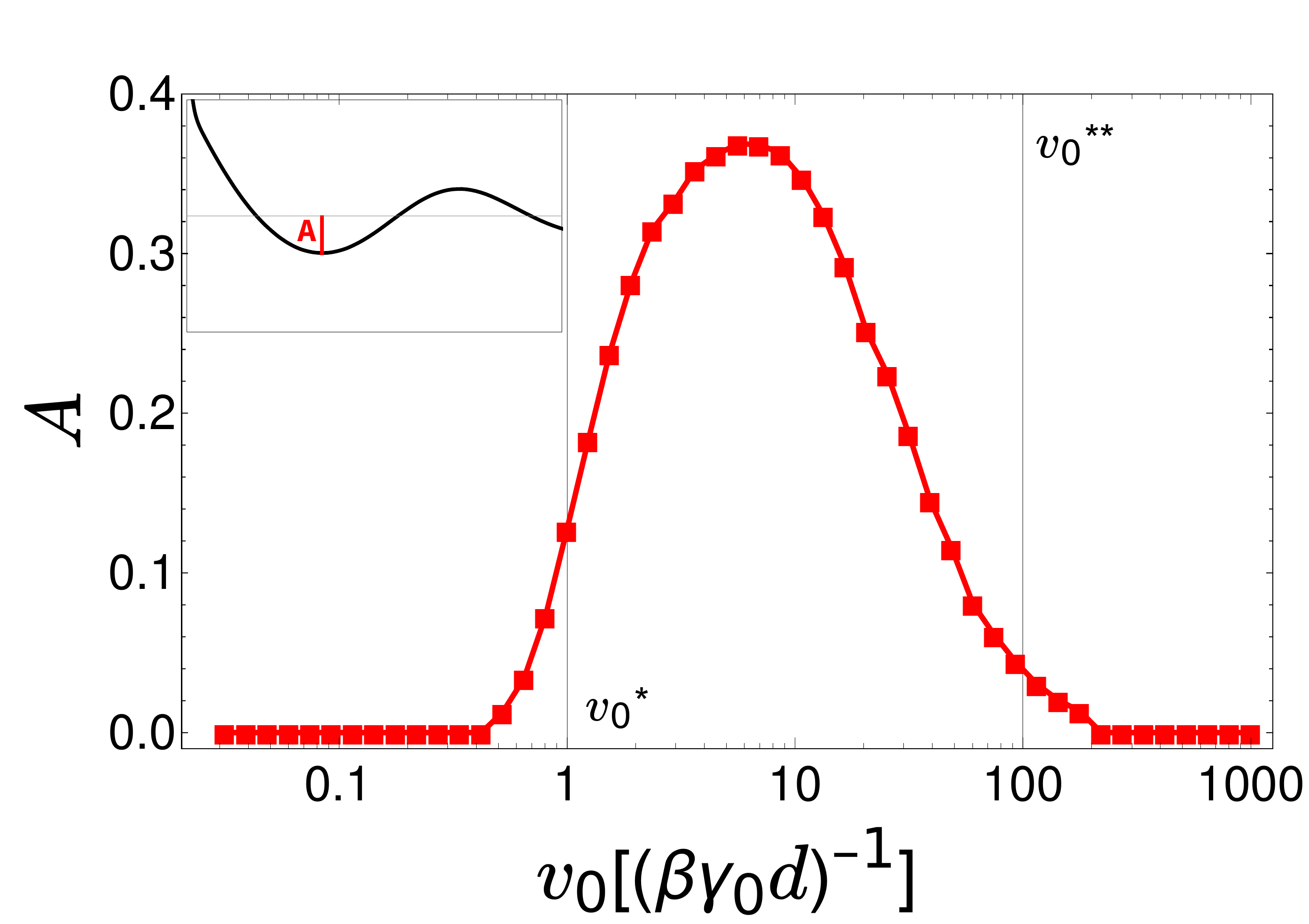}
	\caption{Oscillations amplitude of the correlation curves (shown in Fig.~\ref{fig-MCDs_and_CFs}) vs the driving velocity $v_0$. Inset illustrates how the amplitude is derived from the correlation function.}
    \label{fig-amplitude_plot}	
\end{figure}

In Fig.~\ref{fig-amplitude_plot}, we present the dependence of the amplitude of oscillations (taken from the correlation function  of Fig.~\ref{fig-MCDs_and_CFs}) on dragging velocity $v_{0}$. With increasing $v_{0}$, the amplitude of oscillations, obtained from the curves as shown in the inset sketch, first increases, reaches a maximum value, and then decays back to zero. As expected, for small values of $v_{0}$, i.e. in the linear response regime, the correlation function or the MCD decays monotonically without exhibiting any noticeable oscillations, and therefore the amplitude  is zero. For the given choice of parameters (large value of $\beta V_0$), this corresponds to the situation where tracer and bath particles move (almost) with the same average speeds. Notably, for smaller values of $\beta V_0$, the two particles move at different speeds even in linear response (and show no oscillations), as will be detailed in Sec.~\ref{sec:Pin} below. As we move further away from equilibrium and pass the critical velocity $v_{0}^{\ast}=\frac{2\pi V_0}{\gamma_b d}$, the tracer and bath particles travel with different average speeds. Oscillations appear, and their amplitudes grow. 
The oscillations then reach a maximum at a velocity of around 10 in the units of $(\beta\gamma_0d)^{-1}$, roughly corresponding to the end of shear-thinning regime. The tracer and bath particle then become more an more decoupled, upon crossing the velocity scale of $\kappa d/\gamma_0$ (100 in the units of the graph) and the influence of the bath particle (such as oscillations) disappear. 

\begin{figure}[ht!]
	\centering
	\includegraphics[scale=0.3]{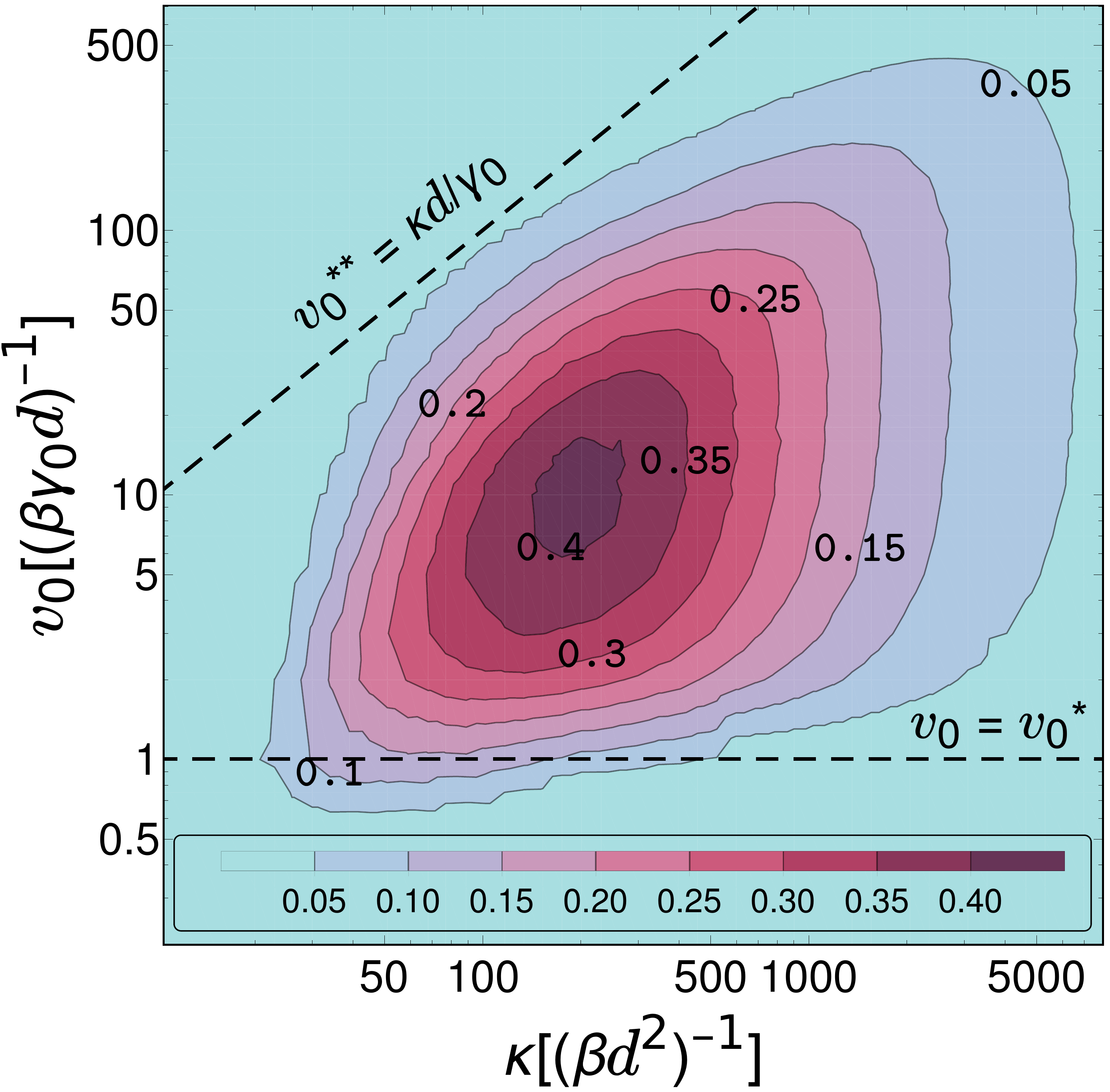}
	\caption{Contour lines of oscillation amplitude as function of $\kappa$ and $v_{0}$ for a barrier height of $\beta V_{0}=4$, in steps of $0.05$. The dashed lines mark two important velocity scales. Below the critical velocity, i.e. $v_0\alt v_0^{\ast}$, the system is close to equilibrium, and no oscillations occur. For $v_0\agt v_0^{**}=\kappa d/\gamma_0$, the frequency of oscillations is large compared to the eigenfrequency of the trapped particle, and the amplitude goes down.}
	\label{fig-ContourPlots}
\end{figure}

In Ref.~\cite{Berner2018Oscillating}, an important observation was made, in experimental data, as well as in a phenomenological Langevin equation: The amplitude of oscillations is a strong function of trap stiffness $\kappa$, so that only a certain range of traps is in resonance with the fluid. Fig.~\ref{fig-ContourPlots} investigates this in the current model by showing the amplitude of oscillations as a function of $\kappa$ and $v_{0}$ keeping all other parameters fixed. Indeed, oscillations appear only for in a finite range of values for $\kappa$ and $v_{0}$, showing a maximum at $v_0\approx 10$ and $\kappa\approx 200$ in the given units. The SPT model is thus in agreement with these previous experimental observations.

In Fig.~\ref{fig-ContourPlots}, we also mark two relevant velocity scales seen in the flow curve of Fig.~\ref{fig:flowcurve}:   The critical velocity $v_0^{\ast}$, below which no oscillations occur, and $v_0^{\ast\ast}= \kappa d/\gamma_0$, above which the frequency of oscillations is large compared to the eigenfrequency of the trapped particle, and the amplitude of oscillations becomes negligibly small.

\subsection{Frequency of Oscillations}
\begin{figure}
    \centering
    \includegraphics[width=0.45\textwidth]{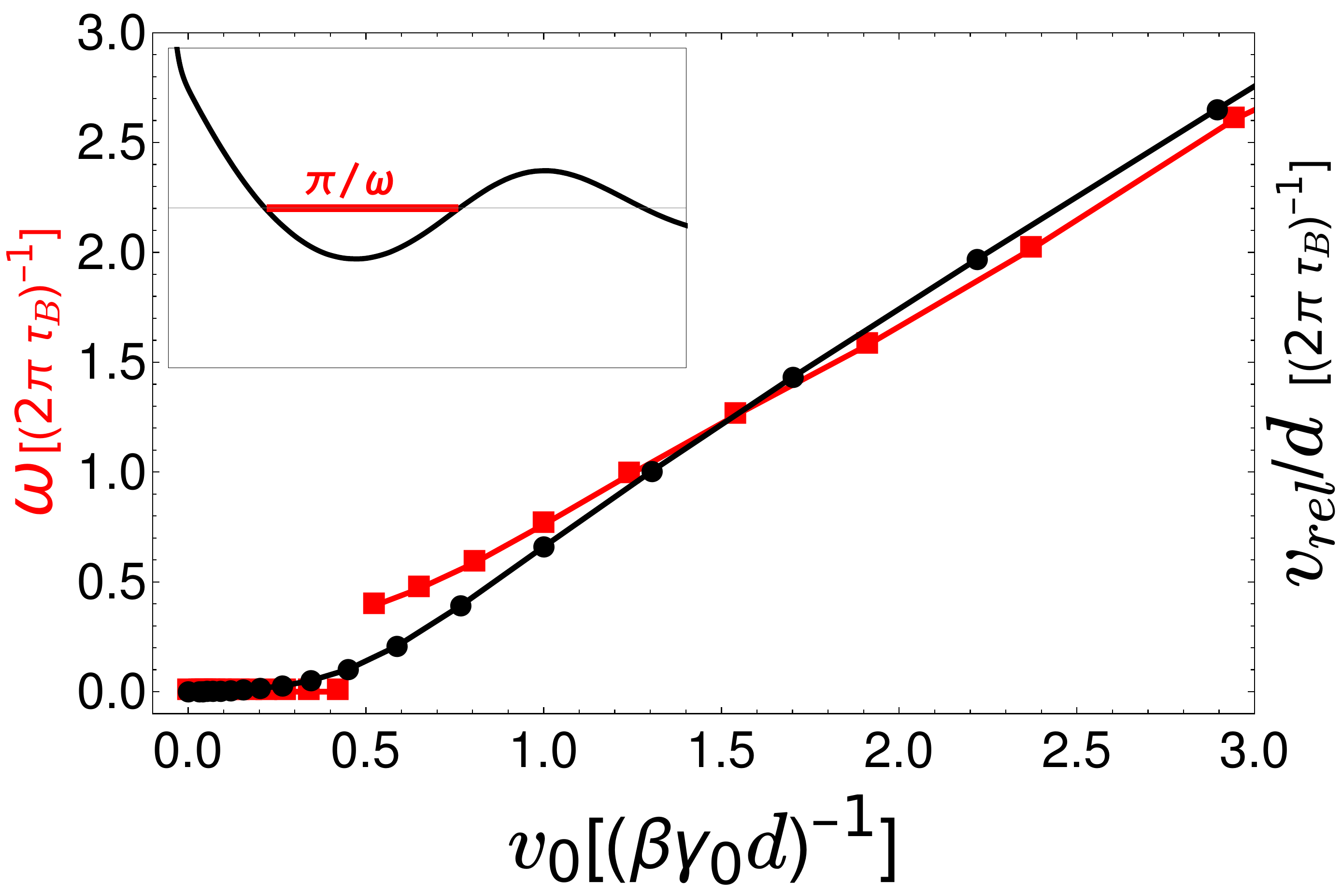}
	\caption{Frequency of the oscillations (red squares) in the correlation curves (shown in Fig.~\ref{fig-MCDs_and_CFs}) versus the driving velocity $v_0$. Inset illustrates how frequency is extracted from the correlation curve. The oscillation frequency is also compared with the average relative velocity computed from the flow curve (black disks). The latter is a smooth function of velocity $v_0$ as illustrated in Eq.~(\ref{freq_from_fc}).}
    \label{fig-frequecy_plot}	
\end{figure}

In Fig.~\ref{fig-frequecy_plot}, it is seen that the frequency of oscillations, obtained from the correlation curve (see inset), increases (almost) linearly with $v_{0}$, which is in agreement with experimental observations, see Fig. 4b) of Ref.~\cite{Berner2018Oscillating}; a strong hint that some of the physics described by this model  indeed has experimental relevance. It is easy to realize that the frequency should be related to the {\it difference} in average velocities of tracer and bath particles in the non-equilibrium steady-state. We thus compare the frequency of oscillations $\omega$, obtained from time dependent correlation functions, to the average relative velocity $\left\langle v_{\textrm{rel}}\right\rangle$ using the relation
\begin{equation}
   \omega = \frac{2\pi}{d}\,\left\langle v_{\textrm{rel}}\right\rangle = \frac{2\pi}{d}\, \left(v_{0} - \left\langle v_{b}\right\rangle \right). \label{freq_from_rel_vel}
\end{equation} 
We used that the average velocity of tracer equals $v_{0}$, whereas the average velocity of bath $\left\langle v_{b}\right\rangle$ is different in general. The result of Eq.~\eqref{freq_from_rel_vel} is shown in Fig.~\ref{fig-frequecy_plot} on the right axis, showing good agreement, except  around $v_0^*$. In that regime it is difficult to obtain the frequency from the correlation function (or MCD) directly, because the amplitude and frequency are both small. Eq.~\eqref{freq_from_rel_vel} is easier to evaluate, and we will use it in Sec.~\ref{sec:Pin} to discuss the behavior at larger $V_0$.

Notably, the frequency, as obtained from Eq.~\eqref{freq_from_rel_vel} is exactly related to the flow curve, via a relation for the total friction force in steady state,
\begin{align}
    \gamma(v_0)v_0=v_0\gamma_0+\left\langle v_{b}\right\rangle\gamma_b.
\end{align}
Solving this for $\left\langle v_{b}\right\rangle$ and plugging into Eq.~\eqref{freq_from_rel_vel}, yields
\begin{equation}
   \omega =  \frac{2\pi}{d}v_0\, \left(1 -\frac{\gamma-\gamma_0}{\gamma_b}\right)   \label{freq_from_fc}.
\end{equation}
This relation shows that $\omega\to \frac{2\pi}{d}v_0$ as $\gamma\rightarrow\gamma_0$ for large $v_0$. The frequency of oscillations is thus a good measure to determine the scale $d$, when comparing the SPT model to experiments \cite{jain2021step}. $\omega$ is minimal for  $v_0\to0$, and it goes  to zero in this limit if the potential $V_{\textrm{int}}$ is large.

\section{Limit of large $\kappa$: Rupture transition}
\label{sec:Pin}
In this section, we discuss the limit of $\kappa\to\infty$, as it allows analytical expressions in some cases.   In this limit the tracer moves with the prescribed velocity $v_0$, and $X\equiv 0$, so that there are naturally no oscillations in this limit. We can however use the definition for frequency of Eq.~\eqref{freq_from_rel_vel} to obtain insights into the behavior of this system. The equations of motion in this limit read
\begin{eqnarray}
	\dot{x}(t) &=& v_{0} \label{kappa_infinite_eq1} \\
	\gamma_{b}\,\dot{q}(t) &=& \frac{2\pi}{d}\,V_{0}\, \sin\left(\frac{2\pi}{d}(x-q)\right) + \eta_{b}(t). \label{kappa_infinite_eq2}
\end{eqnarray}
Using the relative coordinate, $z=q-x$, we can rewrite the above system of equations as
\begin{equation}
\gamma_{b}\,\dot{z}(t) = -\gamma_{b}v_{0} + \frac{2\pi}{d}\,V_{0}\, \sin\left(\frac{2\pi}{d}\,z\right) + \eta_{b}(t). \label{LangevinEqn_RelativeCoordinate}
\end{equation}
Eq.~(\ref{LangevinEqn_RelativeCoordinate}) corresponds to the problem of Brownian motion in a periodic potential, {\it tilted} by a force $-\gamma_b v_0$. This problem arises in several fields of science and has been a subject of significant scientific interest in the literature (see e.g. Ref.~\cite{Risken} and references therein). Thus, in the limit of $\kappa\rightarrow\infty$, the problem of dragging a particle in a viscoelastic fluid (modeled with the SPT model) maps onto the problem of motion of a Brownian particle in a tilted periodic potential (with diffusion coefficient of the bath particle).

The normalized steady-state distribution for the relative coordinate $z$ can be computed analytically~\cite{Risken}, to give
\begin{equation}
	P_{\textrm{st}}(z) = \frac{1}{N}\,e^{-\beta U(z)}\left[ 1 + \frac{\left(e^{\beta\gamma_{b}v_{0}d} -1\right)}{\int_{0}^{d}dz'\,e^{\beta U(z')}}\, \int_{0}^{z}dz'\,e^{\beta U(z')}\right]
\end{equation}
where
\begin{equation}
	U(z) = \gamma_{b}\,v_{0}z - V_{0} \cos\left(\frac{2\pi}{d}z\right)
\end{equation}
where $N$ is the normalization factor such that $\int_{0}^{d}dz\,P_{\textrm{st}}(z)=1$. 
The details of the above derivation are provided in Appendix~\ref{appendix-some_exact_results}. The  steady-state distribution allows to compute the flow curve, and from it, using Eq.~\eqref{freq_from_fc}, the frequency of oscillations (bearing in mind that the oscillations have zero amplitude in the given limit).
For small velocity $v_0$, this yields for the friction $\gamma$
\begin{equation}
\lim\limits_{v_{0}\rightarrow 0}\gamma(v_{0}) =  \gamma_{0} + \gamma_{b}\left[1 - \frac{1}{I_{0}^{2}(\beta V_{0})}\right] 
\label{eq:sv}
\end{equation}
where $I_{0}$ is the modified Bessel function. This formula predicts the dependence of flow curve on $V_0$: For $\beta V_0\ll 1$, the modified Bessel function approaches $1$, and $\gamma$ goes to $\gamma_0$, as expected. For $\beta V_0\gg 1$, the modified Bessel function goes to infinity as $I_{0}(\beta V_{0}) \simeq e^{\beta V_0}\, \frac{1}{\sqrt{2\pi\beta V_{0}}}$ such that 
\begin{equation}
  \lim\limits_{v_{0}\rightarrow 0,\beta V_0\to \infty}  \gamma = \gamma_0 + \gamma_{b} \left[1 - 2\pi\beta V_{0}\,e^{-2\beta V_{0}} \right],
\end{equation}
i.e. in the limit of infinite barrier height $\gamma=\gamma_0 + \gamma_b$, as expected.
Using Eq.~\eqref{freq_from_fc}, the frequency, for $\beta V_0\to0$, is just $\omega=\frac{2\pi}{d} v_0$. For $\beta V_0\to\infty$, we have
\begin{align}
    \lim\limits_{v_{0}\rightarrow 0,\beta V_0\to \infty}\omega = \frac{4\pi^{2}\beta V_{0}}{d} v_{0}\, e^{-2\beta V_{0}}  .
\end{align}
\begin{figure}[ht!]
	\centering
	\includegraphics[scale=0.33]{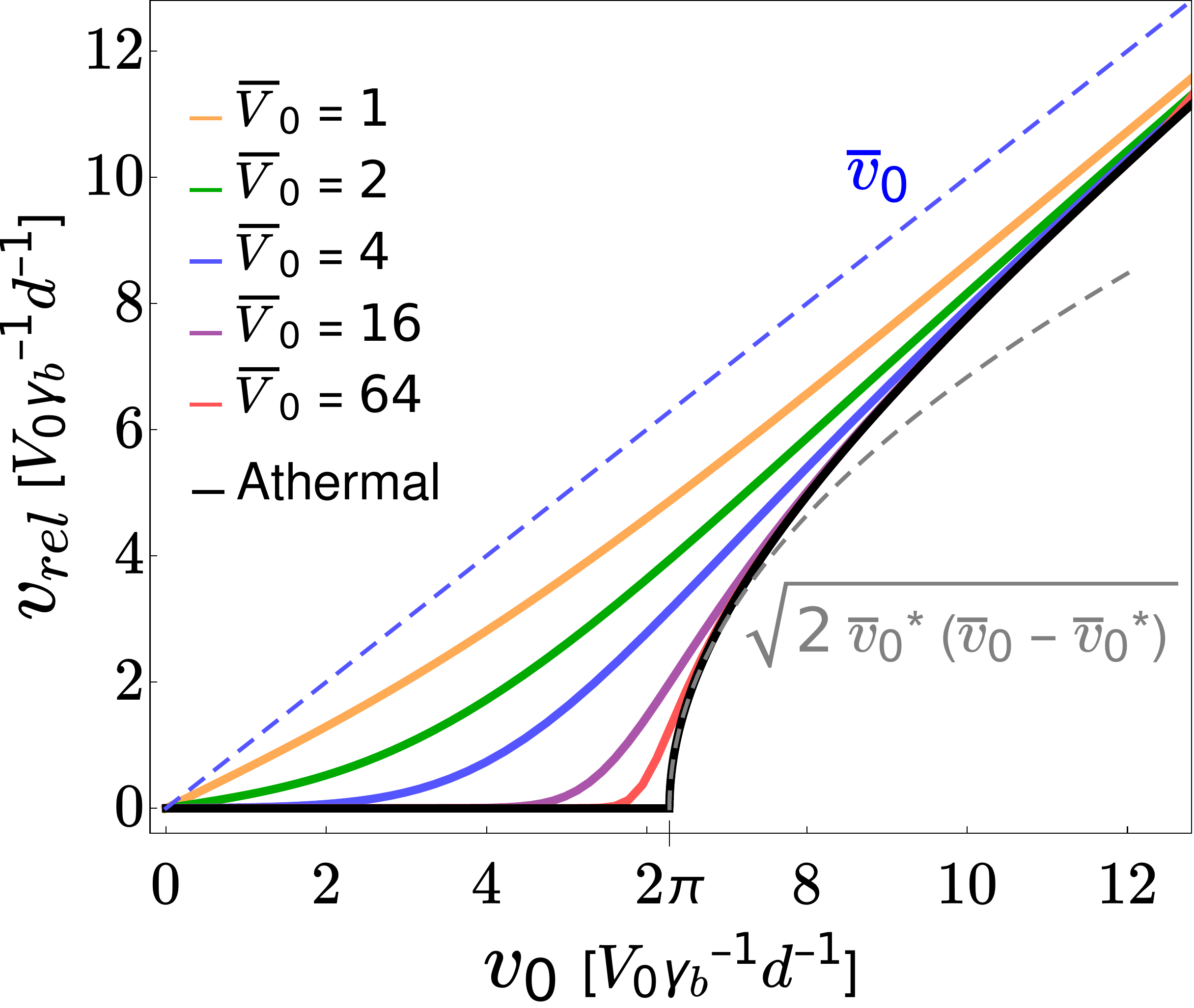}
	\caption{Average relative velocity versus $\bar{v}_{0}$ for different values of $\bar{V}_0$ for the limiting case of infinite trap-stiffness. The black line corresponds to the athermal limit, i.e. $\beta V_{0}\rightarrow\infty$ limit as derived in Eq.~(\ref{average_relative_velocity-athermal_limit}). In the proximity of critical velocity $v_{0}^{\ast}$, relative velocity increases proportional to  $\sqrt{v_0-v_0^*}$ (dashed gray curve) and in the limit of large driving velocity, relative velocity increases linearly with $v_0$ (dashed blue line).}
\label{fig-relative_velocity_infinite_trap}	
\end{figure}
In Fig.~\ref{fig-relative_velocity_infinite_trap}, we plot the relative velocity using the formula~(\ref{av_rel_vel}), as a function of $v_0$, for various values of $V_0$. For small $v_0$, we observe the behavior derived from Eq.~\eqref{eq:sv}, i.e., the slope varies between 0 and $\frac{2\pi}{d}$. For large $v_0\gg v_{0}^{\ast}$, we have $\omega=\frac{2\pi}{d}v_{0}$ for any value of $V_0$. In between however, the curve develops a more and more pronounced behavior as $\beta V_0$ gets large: For $\beta V_0\to\infty$, the following behavior is found
\begin{equation}
    \left\langle v_{\textrm{rel}}\right\rangle = 
    \begin{cases}
    0, \qquad\qquad\qquad\quad\, \text{for $v_{0}\le v_{0}^{\ast}$} \\
    \sqrt{v_{0}^{2}-\frac{4\pi^{2}V^{2}_{0}}{\gamma^{2}_{b}d^{2}}}, \qquad \text{for $v_{0}> v_{0}^{\ast}$}
    \end{cases}
    \label{average_relative_velocity-athermal_limit}
\end{equation}
This is shown as the black line in Fig.~\ref{fig-relative_velocity_infinite_trap}, and it is identical to the athermal limit of zero temperature. It is thus found from  the noise-free solution of equation~(\ref{LangevinEqn_RelativeCoordinate}).

Eq.~\eqref{average_relative_velocity-athermal_limit} displays a \textit{rupture transition}: Tracer and bath particles move together at the same speed for $v_0\leq v_0^*$, and move at different speeds for $v_0> v_0^*$. Near the critical velocity of $v_{0}^{\ast} = \frac{2\pi V_{0}}{\gamma_b d}$, the frequency follows
\begin{equation}
    \left\langle v_{\textrm{rel}}\right\rangle = \sqrt{2v_{0}^{\ast}}\left(v_{0}-v_{0}^{\ast}\right)^{\frac{1}{2}}.
\end{equation}
The rupture transition thus displays a critical exponent of $\frac{1}{2}$, reminiscent of the mean field result for a system close to a second order phase transition~\cite{kardar2007statistical}.

\section{Relation between oscillations and friction}
\label{sec:OscillationsIncreaseFriction}
In this section, we come back to finite values of trap stiffness $\kappa$, investigating the influence of oscillations on friction.

\subsection{Steady driving}
Comparing the amplitude of oscillations in Fig.~\ref{fig-amplitude_plot} to the flow curve, Fig.~\ref{fig:flowcurve}, we note a relation: In the regime where the amplitude is large, i.e, for values of $\bar v_0$ between 1 and 100, the flow curve seems to show a regime of smaller slope. It is a shoulder, or may even be termed as a (very) mild second plateau.

\begin{figure}
\centering
	\includegraphics[width=0.45\textwidth]{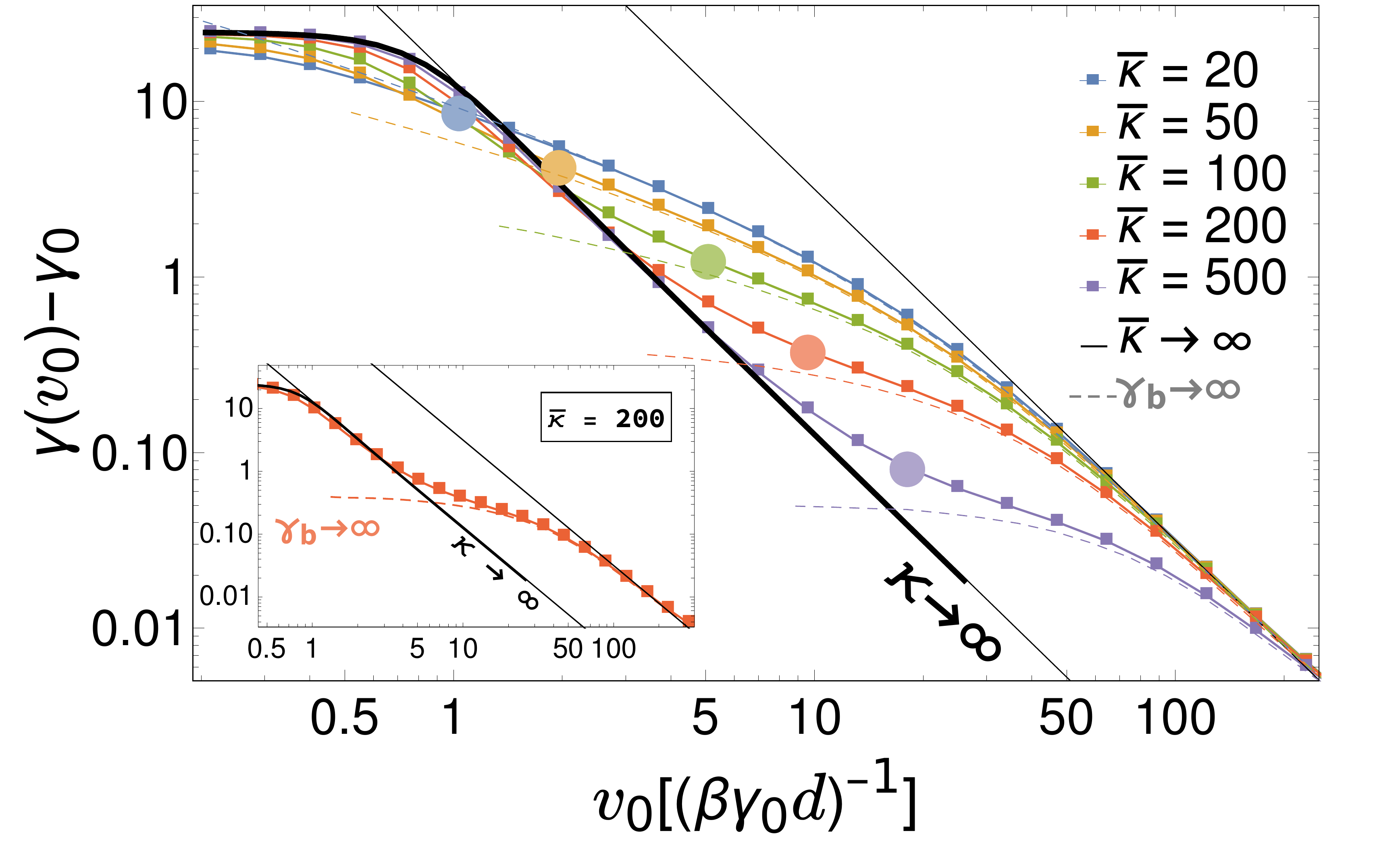}
    \caption{Flow-curve in units of tracer's friction coefficient $\gamma_0$ for different values of trap-stiffness and a barrier height $\beta V_0=4$. The parameters are given in the units discussed in Sec.~\ref{PTModel}. 
    The highlighted point (color circle) on a given flow curve marks the point of largest amplitude observed in the correlation function. The colored dashed lines are the corresponding flow-curves with an infinite bath friction, i.e. $\gamma_b\to\infty$. The thin black lines, left and right, correspond to $0.49\frac{v_{0}^{\ast 2}} {v^{2}_{0}}(\gamma_{0}+\gamma_{b})$ and $0.50\frac{v_{0}^{\ast 2}} {v^{2}_{0}} \frac{\gamma_{b}^{2}}{\gamma_0}$, respectively, where in both cases the numerical prefactor is found by fitting. The oscillations (in the MCD curve) seem to appear in the region set by these two boundaries. The inset shows the curves for a single trap-stiffness $\bar{\kappa}=200$ for better visibility.  
    }
\label{fig:flowcurve_vs_kappa_V0=4}
\end{figure}

Aiming to investigate this further, Fig.~\ref{fig:flowcurve_vs_kappa_V0=4} shows the flow curve for different values of trap stiffness $\kappa$, keeping all other parameters identical to the previous section (see Table~\ref{table:simulation_parameters-set-2}), in particular, $\beta V_{0}=4$. We also subtract $\gamma_0$ and show a logarithmic presentation to highlight the approach of the large velocity regime.

In the linear response regime, the friction $\gamma$ increases with the trap-stiffness $\kappa$, an effect which is well known~\cite{Squires2005-mc,Daldrop2017external, Muller2020properties}.

This trend however is inverted for larger velocities, an effect which can (at least partially) be attributed to oscillations: In the curves, we mark the velocity of maximal amplitude of oscillations (big bullets), showing a remarkable feature: the position of the shoulder in the flow curve seems to coincide with the maximal  amplitude of oscillations. This shows that oscillations may have a strong impact on microrheological friction. One interpretation might be that the excited internal degree of freedom provides additional dissipation, and hence larger friction.

\begin{figure}
\centering
	\includegraphics[width=0.45\textwidth]{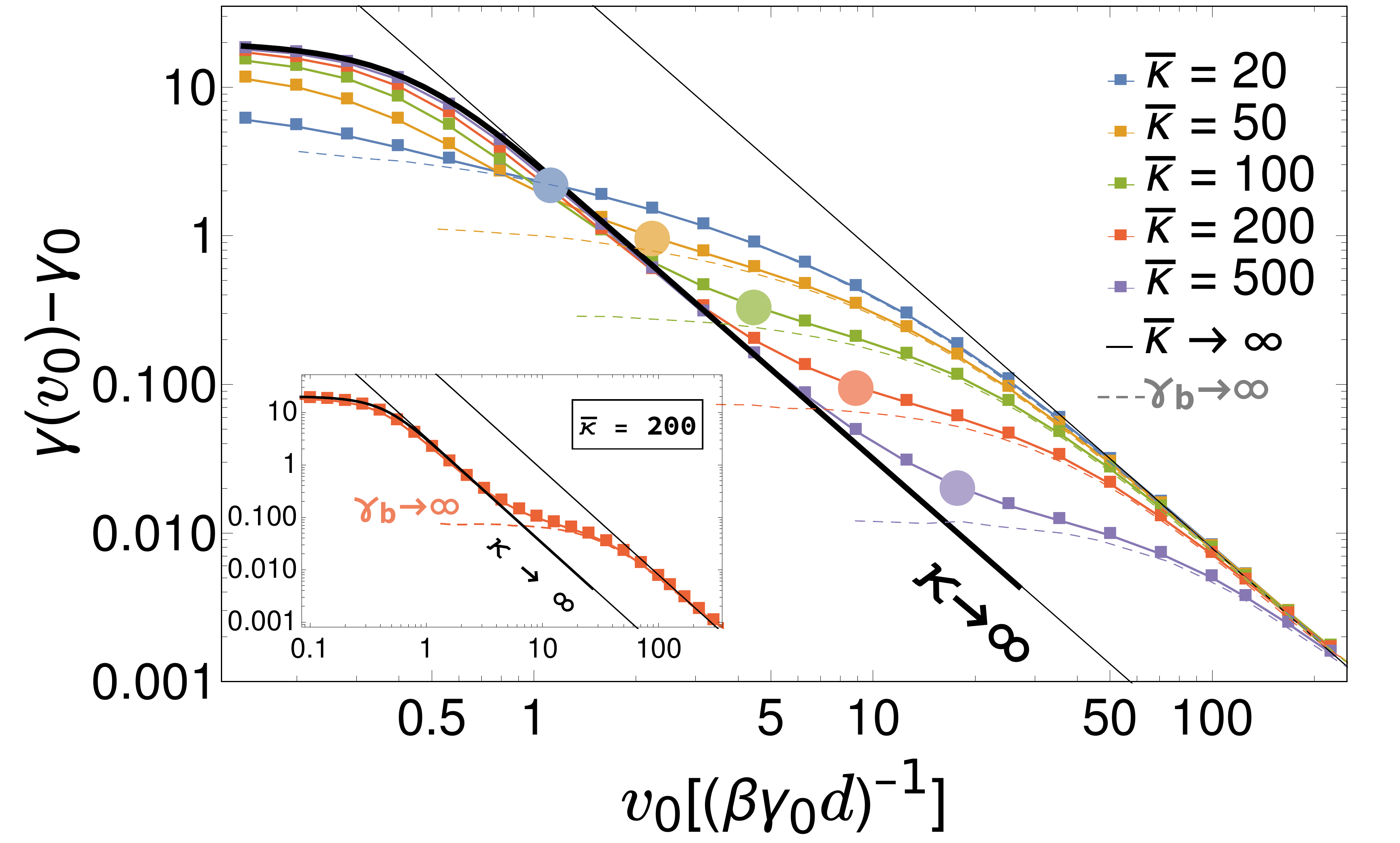}
    \caption{Flow-curve in units of tracer's friction coefficient $\gamma_0$ for different values of trap-stiffness and a barrier height $\beta V_0=2$ (all other parameters identical to Fig.~\ref{fig:flowcurve_vs_kappa_V0=4}). The parameters are given in the units discussed in Sec.~\ref{PTModel}. The highlighted point (color circle) on a given flow curve marks the point of largest amplitude observed in the correlation function. The colored dashed lines are the corresponding flow-curves with an infinite bath friction, i.e. $\gamma_b\to\infty$. The thin black lines, left and right, correspond to $0.48\frac{v_{0}^{\ast 2}} {v^{2}_{0}}(\gamma_{0}+\gamma_{b})$ and $0.50\frac{v_{0}^{\ast 2}}{v^{2}_{0}} \frac{\gamma_{b}^{2}}{\gamma_0}$, respectively, where the prefactors are found by fitting. The oscillations (in the MCD curve) seem to appear in the region set by these two boundaries. The inset shows a particular trap-stiffness, $\bar{\kappa}=200$, for better visibility.  
    }
\label{fig:flowcurve_vs_kappa_V0=2}
\end{figure}

Related to the transition from linear response to the case with oscillations, a remarkable transition towards the behavior of the original overdamped PT model (i.e., $\gamma_b=\infty$) is observed: the results of $\gamma_b=\infty$ are shown in the graph as well, and they agree for large velocities, including partly the regime of oscillations. The flow curve thus shows three  regimes: For $v_0\ll v_0^*=2\pi\frac{V_0}{\gamma_b d}$, we have the linear regime. For $v_0\agt v_0^*$, shear thinning is observed. For $\kappa\to \infty$ (black line), this follows $\gamma\approx \frac{{v_0^*}^2}{v_0^2} (\gamma_0+\gamma_b)$ - a phenomenological relation found from the observation of the power law of $v_0^{-2}$, which is matched with the linear regime at $v_0^*$. Notably, for $\kappa=\infty$, the original PT model shows no friction, i.e., $\gamma=\gamma_0$ for all $v_0$. The $\kappa=\infty$ curve is thus not present in that model (the black line shifts to the left with increasing $\gamma_b$). For finite $\kappa$ the original PT model shows finite friction, and, remarkably,  to the right of the $\kappa=\infty$ curve, the results for finite $\kappa$ quickly approach the $\gamma_b=\infty$ curves. The intersection of the curves of $\kappa=\infty$ and $\gamma_b=\infty$ thus sets another velocity scale, where the behavior of the original PT model is approached. That scale is however difficult to estimate, as the behavior of the original PT model is manifold \cite{muser11a,Popov,muser2020shear}.

Once the original PT model is approached, the flow curve follows the fate set by that model, which shows the mentioned velocity scale of $\kappa d/\gamma_0$, relating the time scale of relaxation in the potential to the frequency of oscillations. Once that scale is passed, the curve reaches the last regime, where a $\kappa$- and $\gamma_b$-independent power law behavior of $v_0^{-2}$ is observed. The independence of $\kappa$ and $\gamma_b$ suggests a scaling of $\gamma\approx\frac{{v_0^*}^2}{v_0^2}\frac{\gamma_b^2}{\gamma_0}$, which is however a phenomenological guess (notably independent of temperature).  

These effects can as well be observed for smaller values of $V_{0}$ as can be seen in Fig.~\ref{fig:flowcurve_vs_kappa_V0=2}, for which $\beta V_{0}=2$. Here, the dependence on $\kappa$ in the linear regime is even stronger: For the smaller barrier height of $\beta V_{0}=2$, thermal fluctuations explore the nonlinear regimes of the interaction potential more strongly, so that this nontrivial effect is enhanced (recall that this effect is absent for a quadratic interaction potential \cite{Muller2020properties}). On the other hand, for $V_0\to 0$, all viscoelastic effects disappear, and so does the dependence on $\kappa$. The value of $\beta V_0=2$ seems to be in the range where this effect is maximized.

For $\beta V_0=2$, the above discussed velocity regimes are also seen, including the two power laws, where the $\kappa\to\infty$ curve again roughly separates the behavior of the original PT model from the stochastic PT model. The phenomenological curve $\gamma\approx\frac{{v_0^*}^2}{v_0^2}\frac{\gamma_b^2}{\gamma_0}$ for the final power law fits here as well.

\subsection{Driving with time-dependent velocity}
The previous subsection displayed a coupling between the friction force and oscillations. Can this coupling be stimulated  by application of modulated driving? To investigate this, we introduce in this subsection a time-dependent driving protocol, choosing a periodic driving of the form,
\begin{equation}
    v_{0}(t) = v_{00} + v_{01}\,\cos(2\pi\omega_{0}t) \label{eq:prot}
\end{equation}
where $\omega_{0}$ is frequency of  driving.
The protocol \eqref{eq:prot} is chosen to test the system's response for specific frequencies, expecting a strong response if $\omega_{0}$ matches the frequency of oscillations.
 
When driving the particle with the protocol \eqref{eq:prot} in steady state, we measure the mean friction force that the particle experiences, via 
\begin{equation}
  \langle F\rangle_t =  \kappa\,\omega_0\int_0^\frac{1}{\omega_0} dt\left\langle  x(t) - \int_{-\infty}^t v_{0}(t')dt'\right\rangle . \label{eq:mf}
\end{equation} 
Averaging over one cycle in steady state yields the time independent results. 

\begin{figure}
	\centering
	\includegraphics[scale=0.3]{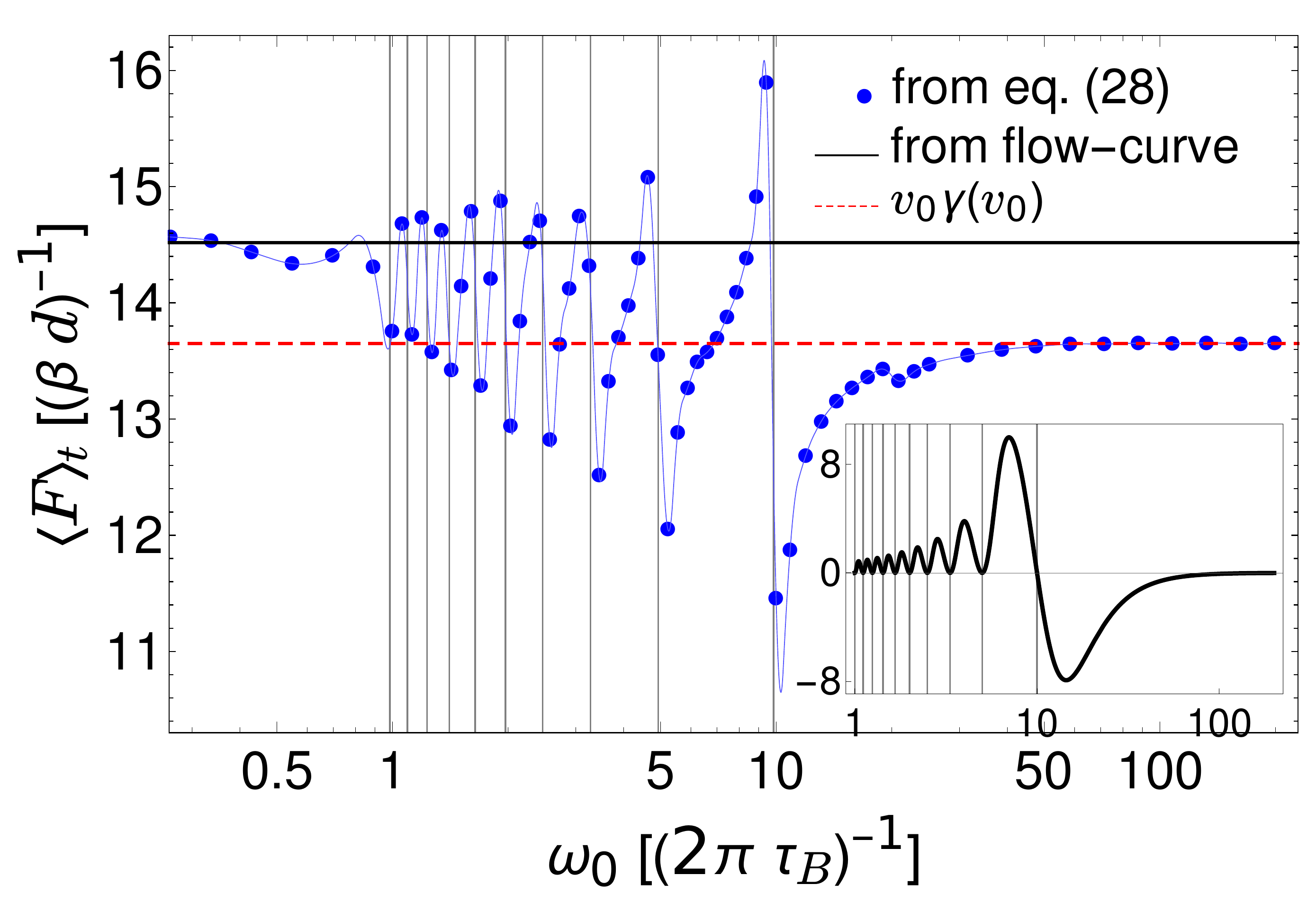}
	\caption{Average friction force versus ${\omega}_{0}$ for the state point with maximal amplitude of oscillations in Fig.~\ref{fig-ContourPlots},  i.e. for $\bar{V}_{0}=4$, $\bar{\kappa}_{0}=200$ and $\bar{v}_{00} =10$. We show the case $\bar{v}_{01}=10$, so that driving velocity modulates between zero and $2v_0$.  The blue line is a guide to the eye. The black line gives the value obtained from the flow-curve, see main text. The red dashed line corresponds to the value at ${v}_{01} =0$. For this case, the frequency of oscillations for steady driving is $\bar{\omega}=9.854$, and the vertical lines at $\bar{\omega}/n$, with integer $n$, mark the \textit{lower} harmonics. Inset gives the result of Eq.~\eqref{deltaF_shear-thinning} (with $\gamma(v_0) v_0$ subtracted) showing a similar oscillating pattern with the zeros occurring at \textit{lower} harmonics $\bar{\omega}_0=10/n$ for integer $n$.}
\label{fig-force_vs_frequency-shear_thinning}	
\end{figure}

Fig.~\ref{fig-force_vs_frequency-shear_thinning} shows the resulting mean force for the state-point corresponding to the maximum amplitude of oscillations in Fig.~\ref{fig-ContourPlots}; a state-point where we expect a pronounced interplay between oscillations and driving. In the figure, we show the mean force as a function of $\omega_0$, choosing $v_{01}=v_{00}$. For the chosen mean velocity of $\bar v_0=10$, the frequency of oscillations is close to $\bar \omega_0=10$ (compare Fig. \ref{fig-frequecy_plot} for actual frequency of oscillations, $\bar{\omega}=9.854$). Indeed, around that value of $\bar \omega_0=10$, the effect of driving on the mean force is strongest as seen in the graph. The graph however shows  more unexpected features, namely, a number of  {\it lower} harmonics, with repeated cycles of larger and smaller mean force. Notably, close to the resonance, the mean force shows a sharp reduction by $28$ \% when going from $\bar{\omega}_0= 10$ to   $\bar{\omega}_0= 10.88$.

\begin{figure}
	\centering
	\includegraphics[scale=0.3]{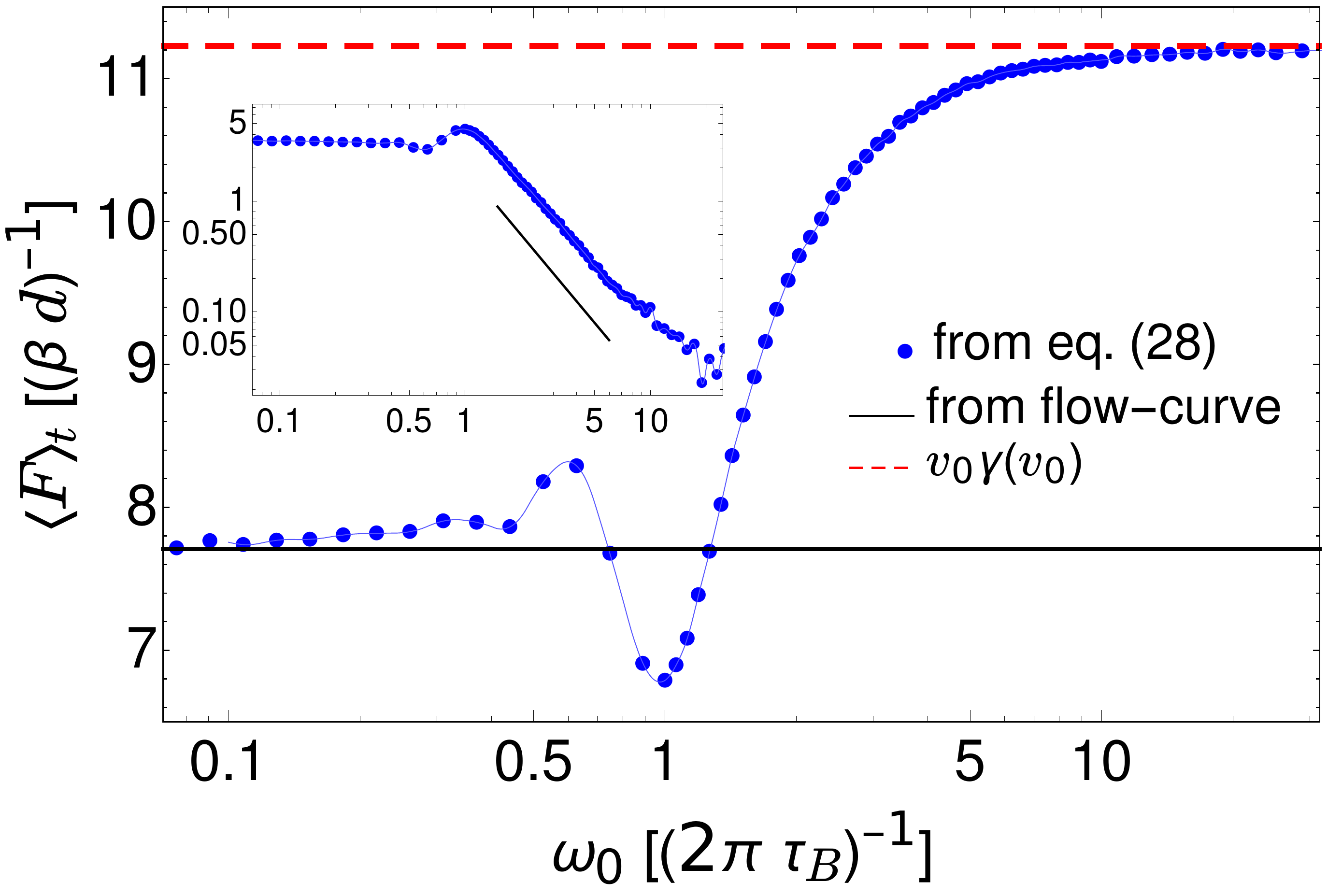}
	\caption{Average friction force versus ${\omega}_{0}$ near the linear response regime with parameters $\bar{V}_{0}=4$, $\bar{\kappa}_{0}=200$ and $\bar{v}_{00} =1$. We show the case $\bar{v}_{01}=1$, so that driving velocity modulates between zero and $2v_0$. Blue line is a guide to the eye. The black line refers to the small frequency limit, obtained from the flow-curve, see main text. The red dashed line corresponds to the flow curve value, i.e.,  viscosity corresponding to $\bar{v}_{0} =1$. The inset shows the log-log plot of $\langle F\rangle_t- \gamma(v_0)v_0$ vs $\omega_0$ of the same data, demonstrating an asymptotic power-law scaling in agreement with Eq.~\eqref{deltaF_linear_response} (bar shows the power law $\omega_0^{-2}$).}
\label{fig-force_vs_frequency-linear_response}	
\end{figure}

For small frequency, $\bar\omega_0\ll 1$, for each part of the driving cycle, the system is in the steady state  corresponding to the given driving velocity. In this limit, the mean force can thus be found from the flow curve in Fig.~\ref{fig:flowcurve_vs_kappa_V0=4}, integrated with the corresponding time dependent friction over one cycle,
\begin{equation}
  \langle F\rangle_t = \omega_0 \int_{0}^{1/\omega_0}dt\,\gamma(v_0(t))\,v_0(t), \label{eq:mf-flowcurve}
\end{equation}
where $\gamma(v_0)$ is the flow curve, as used above. This result is shown as a solid black line in Fig.~\ref{fig-force_vs_frequency-linear_response}, and it is indeed approached by the data points. In the opposite limit,  $\bar\omega_0\gg 1$, the time dependent part of driving has no effect on the mean force. The mean force thus approaches the result of $v_{01}=0$, i.e., the steadily driven case, which is shown as the dashed red curve.

Fig.~\ref{fig-force_vs_frequency-linear_response} shows the same quantity for an average velocity near the linear response regime of flow-curve, $\bar v_0=1$, i.e. when the driving velocity is close to the critical value. In this limit, the amplitude of oscillations is smaller and we see that the mean friction force changes almost monotonically from its small frequency limit (solid black line) to large frequency limit (dashed red line).

The behavior observed in  Figs.~\ref{fig-force_vs_frequency-shear_thinning} and \ref{fig-force_vs_frequency-linear_response} can partly be understood analytically in the limit $\kappa\to\infty$, see Sec.~\ref{sec:Pin} above, and Appendix~\ref{appendix-athermal_limit}. Using additionally the athermal limit, $\beta V_0\to\infty$, we can obtain the response to modulated driving perturbatively, i.e, to linear order in $v_{01}$. For $v_{00}\gg v_0^*$, which is the case in Fig.~\ref{fig-force_vs_frequency-shear_thinning}, we find
\begin{equation}
    \langle \Delta F\rangle_t =
    \frac{\gamma_{b}v_{01}v^{\ast}_{0}\omega_{0}d} {\pi\left( v^{2}_{00}-\omega^{2}_{0}d^2\right)}\,\sin^{2}\left(\frac{\pi v_{00}}{\omega_0 d}\right)\, e^{-\frac{v^{\ast}_{0}}{v_{00}} \sin\left(\frac{2\pi v_{00}}{\omega_{0}d} \right)},
     \label{deltaF_shear-thinning}
\end{equation}
where $\left\langle \Delta F\right\rangle_t$ is the change in the mean friction force with respect to the unperturbed state. This function becomes zero exactly at the \textit{lower} harmonics, i.e. $\omega_0 = \frac{1}{n}\frac{v_{00}}{d}$ as shown in the inset, and captures the observed oscillations in the mean friction force.

Near the critical velocity (see Appendix~\ref{appendix-athermal_limit}) this perturbative calculation yields
\begin{equation}
    \langle \Delta F\rangle_t = \frac{\gamma_b v_{01}v^{\ast 2}_{0}}{\omega_{0}^{2}d^2 + 2\pi v^{\ast}_{0} \omega_{0}d + \frac{1}{2}(2\pi v^{\ast}_{0})^2}. \label{deltaF_linear_response}
\end{equation}
This equation goes monotonically between the two limiting regimes of small and large $\omega_0$. For large $\omega_0$, the value of $v_{0}\gamma(v_0)$ is approached as $\omega_{0}^{-2}$. The simulation data displays the same power-law behavior as shown in the inset of Fig.~\ref{fig-force_vs_frequency-linear_response}.  

\section{Conclusion}
\label{sec:conclusion}
We studied the micro-rheological behavior of a simple, yet rich model: Two particles coupled via a periodic potential. This model, which is an extension of the famous Prandtl Tomlinson model for dry friction, reproduces the experimental observations of shear thinning as well as position oscillations of a Brownian particle in a micellar suspension \cite{Berner2018Oscillating, jain2021step}:   The amplitude of oscillations grows with driving velocity (for small velocities), and their frequency is (in good approximation) linear in the driving velocity. For larger velocities, the amplitude of oscillations decreases in the model, a prediction not yet tested experimentally.

We observe that the flow curves, i.e., the micro-viscosity as a function of driving velocity, shows, additionally to shear-thinning, a 'shoulder', which appears tightly connected to the amplitude of oscillations: The shoulder is most pronounced when the amplitude is maximal. Aiming to further exploit this coupling between oscillations and micro-viscosity, we investigate time-modulated driving velocities. We find a pronounced resonance behavior concerning the driving frequency and the frequency of oscillations, so that the resulting mean friction force can be larger or smaller than for the steady driven case. Additionally, a large number of lower harmonics are observed.  

The limit of infinitely stiff trapping potential is notable for two reasons: It  can by treated analytically, and it can be mapped on the well studied problem of diffusion in a tilted potential. For infinite barrier height, a non-analytic behavior of oscillation frequency is found, where frequency is zero (exponentially small) below a critical velocity, and grows as a square root for velocities larger than the critical one. This limit also allows to understand the mentioned resonance behavior and the appearance of lower harmonics analytically. 

Finally, the limit of infinite bath friction, in which case the original PT model is approached, seems to be generally approached for large velocities, when the bath particle hardly moves. The flow curve thus (roughly) crosses over from $\kappa\to\infty$ behavior to $\gamma\to\infty$ behavior as a function of velocity.    Or, saying it differently, from stochastic PT behavior at small velocities to original PT behavior at larger velocities.     

A number of predictions of this manuscript have not been tested experimentally, but appear accessible: Time modulated driving can generally be achieved. The limit  of infinite trap stiffness may be achieved experimentally using feedback loops. It will be exciting to see how far the PT behavior, including critical behavior and instabilities, carries over to experiments in fluids.

\section{Acknowledgements}
The authors thank Boris M\"{u}ller, Johannes Berner, and Clemens Bechinger for their helpful discussions. RJ and MK acknowledge support by the G\"{o}ttingen Campus QPlus program as well as Deutsche Forschungsgemeinschaft via SFB 1432 (Project C05). FG acknowledges support by the Alexander von Humboldt foundation. 


\appendix

\section{Simulation parameters}
\label{parameters}
In this appendix, we provide the simulation parameters used for generating the figures in the main text. 

\subsection{For Sec.~\ref{sec:SimulationsAndResults} and Sec.~\ref{sec:FrequencyAndOscillations}}
To produce figures~\ref{fig:flowcurve} and~\ref{fig-MCDs_and_CFs} in Sec.~\ref{sec:SimulationsAndResults}, and figures~\ref{fig-amplitude_plot} and~\ref{fig-frequecy_plot} in Sec.~\ref{sec:FrequencyAndOscillations}, the parameters of Table~\ref{table:simulation_parameters-set-1} were used with a fixed trap-stiffness, i.e. $\bar{\kappa}=100$. On the other hand, the trap-stiffness value is changed to produce Fig.~\ref{fig-ContourPlots} while keeping other simulation parameters same as in Table ~\ref{table:simulation_parameters-set-1} below. 
\begin{table}[ht!]
	\centering 
	\begin{tabular}{|c|c|c|c|c|}
		\hline
		\hline
		\,$\kappa\, [(\beta d^2)^{-1}]$ \, &  \,$dt\, [\tau_{B}]$ \, &  \,$t_{\textrm{eq}}\, [\tau_{B}]$\, &  \,$\gamma_{b}\, [\gamma_{0}]$\, &  \,$\beta V_{0}$\, \\
		\hline
		$100$  &  $10^{-3.5}$  &  $1000$ & $25.0$ & $4.0$ \\
		\hline
		\hline
	\end{tabular}
\caption{Parameters used for the simulation in Sec.~\ref{sec:SimulationsAndResults} and Sec.~\ref{sec:FrequencyAndOscillations}.}
\label{table:simulation_parameters-set-1}
\end{table}
\subsection{For Sec.~\ref{sec:OscillationsIncreaseFriction}}
Fig.~\ref{fig:flowcurve_vs_kappa_V0=4} in Sec.~\ref{sec:OscillationsIncreaseFriction} is produced using the parameters of Table~\ref{table:simulation_parameters-set-2} for a barrier height $\beta V_0=4$. 
\begin{table}[ht!]
	\centering 
	\begin{tabular}{|c|c|c|c|}
		\hline
		\hline
		    \,$\kappa\, [(\beta d^2)^{-1}]$ \, &  \,$dt\, [\tau_{B}]$ \, &  \,$t_{\textrm{eq}}\, [\tau_{B}]$\, &  \,$\gamma_{b}\, [\gamma_{0}]$\, \\
		\hline
		    $20$   &  $10^{-3}$  &  $1000$ & $25.0$  \\
		\hline
			$50$   &  $10^{-3}$  &  $1000$ & $25.0$  \\
		\hline
			$100$  &  $10^{-3}$  &  $1000$ & $25.0$  \\
		\hline
			$200$  &  $10^{-3}$  &  $1000$ & $25.0$  \\
		\hline
			$500$  &  $10^{-3}$  &  $1000$ & $25.0$  \\
		\hline
		\hline
	\end{tabular}
\caption{Parameters used to generate figures~\ref{fig:flowcurve_vs_kappa_V0=4} and~\ref{fig:flowcurve_vs_kappa_V0=2} in Sec.\ref{sec:OscillationsIncreaseFriction}.}
\label{table:simulation_parameters-set-2}
\end{table}\\
To produce Fig.~\ref{fig:flowcurve_vs_kappa_V0=2}, we have used the same parameters as in Table~\ref{table:simulation_parameters-set-2} above but with a smaller barrier height $\beta V_0=2$.
\\
For figures~\ref{fig-force_vs_frequency-shear_thinning} and~\ref{fig-force_vs_frequency-linear_response} we have used the simulation parameters of Table~\ref{table:simulation_parameters-set-3} below.
\begin{table}[ht!]
	\centering 
	\begin{tabular}{|c|c|c|}
		\hline
		\hline
    		{}  & Fig.~\ref{fig-force_vs_frequency-shear_thinning} & Fig.\ref{fig-force_vs_frequency-linear_response} \\
    	\hline	
		    \,$\kappa\, [(\beta d^2)^{-1}]$ \, & $200$ & $200$ \\
		\hline
		    \,$\beta V_{0}$\, & $4.0$ & $4.0$ \\  
		\hline
		    \,$\gamma_{b}\, [\gamma_{0}]$\, & $25.0$ & $25.0$ \\
		\hline
		    \,$v_{00}\, [(\beta\gamma_{0}d)^{-1}]$\, & $10.0$ & $2.0$ \\
        \hline
		    \,$v_{01}\, [(\beta\gamma_{0}d)^{-1}]$\, & $10.0$ & $2.0$ \\
        \hline
		    \,$dt\, [\tau_{B}]$ \, & $10^{-3}$ & $10^{-3}$ \\
        \hline
		    \,$t_{\textrm{eq}}\, [\tau_{B}]$\, & $1000$ & $1000$  \\
		\hline
		\hline
	\end{tabular}
\caption{Parameters used in the simulation to produce figures~\ref{fig-force_vs_frequency-shear_thinning} and~\ref{fig-force_vs_frequency-linear_response}.}
\label{table:simulation_parameters-set-3}
\end{table}

\section{Flow-curve and the average relative velocity}
\label{flowcurve_and_mean_relative_velocity}
The average friction force in the steady-state for a given drag-velocity $v_0$ is nothing but the sum of frictional forces acting on the tracer and bath particles. Thus, the flow-curve can be written as
\begin{equation}
    \gamma(v_{0}) = \frac{1}{v_{0}}\,(\gamma_{0}v_{0}+\gamma_{b}\left\langle v_{b}\right\rangle)
\end{equation}
where $\gamma_0$ and $\gamma_b$ are the friction coefficients, and $v_{0}$ and $\left\langle v_{b}\right\rangle$ are the average velocities in the steady-state for the tracer and bath particles, respectively. Expressing $\left\langle v_{b}\right\rangle$ in terms of the average relative velocity, i.e. $\left\langle v_{b}\right\rangle = v_{0} - \left\langle v_{\textrm{rel}}\right\rangle$, the above expression could be written as 
\begin{eqnarray}
    \gamma(v_{0}) &=& \frac{1}{v_{0}}\,(\left(\gamma_{0}+\gamma_{b}\right)v_{0} - \gamma_{b}\left\langle v_{\textrm{rel}}\right\rangle)   \nonumber\\
    &=& \left(\gamma_{0}+\gamma_{b}\right) - \frac{\left\langle v_{\textrm{rel}}\right\rangle}{v_{0}}\,\gamma_{b}.
\end{eqnarray}
Using the above formula, we can thus compute the average relative velocity (thus, the frequency of oscillations) directly from the flow-curve. 

\section{Analytical treatment of the limit of infinite trap-stiffness ($\mathbf{\kappa\rightarrow\infty}$)}
\label{appendix-some_exact_results}

\subsection{Steady-state probability distribution}
We start by writing Eq.~\ref{LangevinEqn_RelativeCoordinate}) in the standard form
\begin{equation}
\gamma_{b}\,\dot{z}(t) = -U'(z)+ \eta_{b}(t), \label{Langevin_equation_in_z}
\end{equation}
where the total potential $U$ is given as
\begin{equation}
U(z) = \gamma_{b}\,v_{0}\,z +  V_{\textrm{int}}(z). \label{potential_z}
\end{equation}
From the Langevin equation~(\ref{Langevin_equation_in_z}), we write the standard Fokker-Planck equation~\cite{Risken} 
\begin{equation}
\frac{\partial P}{\partial t} = \frac{1}{\beta\gamma_{b}}\,\left[ \frac{\partial^{2}}{\partial z^{2}} + \beta\,\frac{\partial}{\partial z} U'(z) \right] P(z,t) = -\frac{\partial S}{\partial z}  \label{FP_eqn-z}
\end{equation}
where $\beta=(k_{B}T)^{-1}$ and the probability current $S$ is given as
\begin{equation}
S(z,t) = - \frac{1}{\beta\gamma_{b}}\left[\frac{\partial P}{\partial z} + \beta\, U'(z)\,P(z,t)\right]. \label{prob_current-z}
\end{equation}\\
In the steady state, i.e. for $t\rightarrow\infty$ limit, the probability current $S$ is just a constant. Thus, to compute the steady-state distribution $P_{\textrm{st}}$, we need to solve the following ordinary differential equation
$$ \frac{dP_{\textrm{st}}}{dz} + \beta\,U'(z)P_{\textrm{st}}(z) = -\beta\gamma_{b}\,S .$$
This can be immediately solved to give the steady-state probability distribution
\begin{equation}
P_{\textrm{st}}(z) = N^{-1}\,e^{-\beta U(z)} - \beta\gamma_{b}\,S\, e^{-\beta U(z)}\, \int_{0}^{z} dz'\,e^{\beta U(z')}, \label{staedy-state_soln-z}
\end{equation}
where $N$ and $S$ are constants, which can be fixed by using the periodicity and normalization condition. It already follows from Eq.~(\ref{staedy-state_soln-z}) that the distribution $P_{\textrm{st}}(z)$ must be periodic with the only requirement that $P_{\textrm{st}}(z)$ is bounded for large enough $z$~\cite{Risken}. Thus, we have
\begin{eqnarray}
P_{\textrm{st}}(z+nd) &=& P_{\textrm{st}}(z),
\end{eqnarray}
where $n$ is an integer. With this, the steady-state distribution for $0\leq z < d$ is given as
\begin{equation}
P_{\textrm{st}}(z) = N^{-1}\,e^{-\beta U(z)}\left[ 1 + \frac{\left(e^{\beta\gamma_{b}v_{0}d} -1\right)}{\int_{0}^{d}dz'\,e^{\beta U(z')}}\, \int_{0}^{z}dz'\,e^{\beta U(z')}\right] \label{normalized_pdf}
\end{equation}
where $N$ is the normalization factor such that $\int_{0}^{d}dz\, P_{\textrm{st}}(z) =1$, i.e.
\begin{equation}
N = \int_{0}^{d}dz\, e^{-\beta U(z)}\left[ 1 + \frac{\left(e^{\beta\gamma_{b}v_{0}d} -1\right)}{\int_{0}^{d}dz'\,e^{\beta U(z')}}\,\int_{0}^{z}dz'\,e^{\beta U(z')}\right]. \label{normalization_factor}
\end{equation}

\subsection{Flow curve}
The flow curve, in the steady-state, is defined as 
\begin{equation}
\gamma(v_{0}) = \lim\limits_{t\rightarrow\infty} \frac{|\left\langle F_{d}(t)\right\rangle |}{v_{0}} ,
\end{equation} 
where $F_{d}(t)$ is the total drag force acting on the system which in turn can be identified as
\begin{equation}
F_{d}(t) = \gamma_{0}\,\dot{x} + \gamma_{b}\,\dot{q}(t) = \left(\gamma_{0} + \gamma_{b}\right)v_{0} - U'(z). \label{friction_force2}
\end{equation}
Thus, the flow curve becomes 
\begin{equation}
\gamma(v_{0}) = \lim\limits_{t\rightarrow\infty} \frac{|\left\langle F_{d}(t)\right\rangle |}{v_{0}} = \left(\gamma_{0} + \gamma_{b}\right) - \frac{\left\langle U'(z) \right\rangle }{v_{0}} \label{def_flowcurve}
\end{equation} 
where the average of a function $f(z)$ in the steady-state is defined as
$$ \left\langle f(z)\right\rangle = \int_{0}^{d}dz\, f(z)\,P_{\textrm{st}}(z). $$
Using the steady state distribution~(\ref{normalized_pdf}), the flow curve can be computed to give
\begin{widetext}
\begin{equation}
\gamma(v_{0}) =  \left(\gamma_{0}+\gamma_{b}\right) - \frac{d}{\beta v_{0}}\,\frac{\left(e^{\beta\gamma_{b}v_{0}d} -1\right)}{\int_{0}^{d}dz\,e^{-\beta U(z)}\int_{0}^{d}dz'\,e^{\beta U(z')} + \left(e^{\beta\gamma_{b}v_{0}d} -1\right)\int_{0}^{d}dz\,e^{-\beta U(z)} \int_{0}^{z}dz'\,e^{\beta U(z')}}.
\end{equation}
\end{widetext}
One could also calculate the $v_{0}\rightarrow 0$ limit exactly for the flow curve. This limit is given as
\begin{equation}
\lim\limits_{v_{0}\rightarrow 0}\gamma(v_{0}) = \gamma_{0} + \gamma_{b}\left[1 - \frac{1}{I_{0}^{2}(V_{0})}\right] 
\end{equation}
where $I_{0}$ is the modified Bessel function.

\subsection{Average relative velocity}
The relative velocity of the tracer with respect to the bath particle is defined as
\begin{equation}
  v_{\textrm{rel}} =  \dot{x}(t) - \dot{q}(t) = -\dot{z}(t) 
\end{equation}
In the steady state, the average relative velocity from the Langevin equation~(\ref{Langevin_equation_in_z}) becomes
\begin{eqnarray}
  \left\langle v_{\textrm{rel}} \right\rangle = \frac{1}{\gamma_{b}}\, \left\langle U'(z) \right\rangle.
  \label{def_relative_velocity}
\end{eqnarray}
It may be noticed, from equations~(\ref{def_flowcurve}) and~(\ref{def_relative_velocity}), that the flow-curve and the average relative velocity in the steady-state are inter-dependent quantities. The relation between these two is given by
\begin{equation}
    \left\langle v_{\textrm{rel}} \right\rangle = \frac{1}{\gamma_{b}} \left[(\gamma_{0}+\gamma_{b})v_{0} - \gamma(v_{0})v_{0}\right].
\end{equation}
\begin{widetext}
Following the steps used in the computation of flow curve, the average relative velocity is given as
\begin{eqnarray}
\left\langle v_{\textrm{rel}} \right\rangle = \frac{d}{\beta \gamma_{b}}\, \frac{\left(e^{\beta\gamma_{b}v_{0}d} -1\right)}{\int_{0}^{d}dz\,e^{-\beta U(z)}\int_{0}^{d}dz'\,e^{\beta U(z')} + \left(e^{\beta\gamma_{b}v_{0}d} -1\right)\int_{0}^{d}dz\,e^{-\beta U(z)} \int_{0}^{z}dz'\,e^{\beta U(z')}}. \label{av_rel_vel}
\end{eqnarray} 
\end{widetext}

\section{Athermal limit}
\label{appendix-athermal_limit}
In the limit $\kappa\rightarrow\infty$, the \textit{noise-free} equations of motion for the PT model are given by
\begin{eqnarray}
	\dot{x} &=& v_{0} \label{PT_noisefree-eqn1}\\  
	\gamma_{b}\,\dot{q} &=& \frac{2\pi}{d} V_{0} \sin\left(\frac{2\pi}{d}(x-q)\right). \label{PT_noisefree-eqn2}
\end{eqnarray}
For the relative coordinate $z=q-x$, the equation becomes
\begin{eqnarray}
	\gamma_{b}\,\dot{z} &=& -\gamma_{b}v_{0} -\frac{2\pi}{d} V_{0} \sin\left(\frac{2\pi}{d} z\right). \label{PT_noise-free}
\end{eqnarray}

\subsection{Steady driving}
When $v_{0}$ is constant, the solution Eq. \eqref{PT_noise-free} with the initial conditions, $x(t=0)=0$ and $q(t=0)=0$ is
\begin{equation}
    z(t) = -\frac{d}{\pi}\, \tan^{-1}\left[ \frac{v_{0} \tan\left(\frac{\alpha\pi t} {d} \right)}{\alpha + \frac{2\pi V_{0}} {\gamma_{b}d} \tan\left(\frac{\alpha\pi t} {d}\right)}\right] \label{solution_noise-free}
\end{equation}
with $\alpha= \sqrt{v_{0}^{2}-\frac{4\pi^{2}V^{2}_{0}}{\gamma^{2}_{b}d^{2}}}$. The time-dependent relative velocity, $v_{\textrm{rel}}(t)=-\dot{z}(t)$ for the noise-free case is then computed as
\begin{equation}
    v_{\textrm{rel}}(t) = \frac{\alpha^{2}v_{0}}{v_{0}^{2}-\frac{4\pi^{2}V^{2}_{0}}{\gamma_{b}^{2}d^{2}}\cos\left(\frac{2\pi\alpha t}{d}\right) + \frac{2\pi V_{0}} {\gamma_{b}d}\alpha\sin\left(\frac{2\pi\alpha t}{d}\right)}.
\end{equation}
\begin{figure} [ht!]
	\centering
	\includegraphics[scale=0.27]{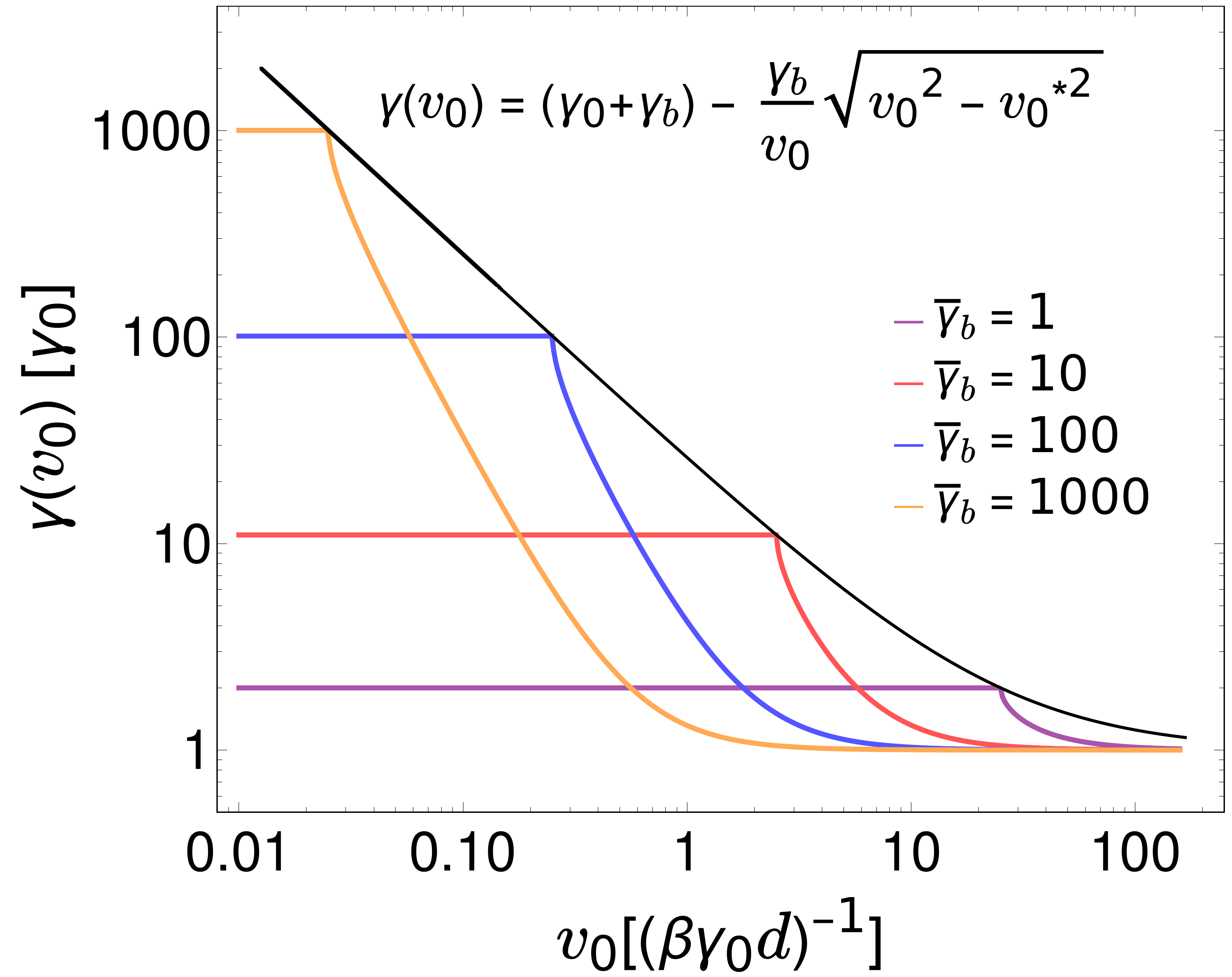}
	\caption{Flow-curve in the athermal limit for different values of $\bar{\gamma}_b$. The black line, joining the turning points, captures the analytical form (also shown in the plot) of the flow-curve. The flow-curve approaches towards its classical analogue in the limit of $\bar{\gamma}_b\rightarrow\infty$. }
	\label{fig-flowcurve_athermal}
\end{figure}\\
For $\alpha<0$, i.e., $v_{0}<\frac{2\pi V_{0}}{\gamma_{b}d}$, $ v_{\textrm{rel}}(t)$ decays to zero with $t$, therefore, the average relative velocity $\left\langle v_{\textrm{rel}}\right\rangle$ for driving speeds smaller than $\frac{2\pi V_{0}} {\gamma_{b}d}$ is zero. For $v_{0}>\frac{2\pi V_{0}} {\gamma_{b}d}$, the quantity $v_{\textrm{rel}}(t)$ oscillates in time with a frequency $\omega=\frac{2\pi\alpha}{d}$ and the average relative velocity $\left\langle v_{\textrm{rel}}\right\rangle$ is finite given by
\begin{equation}
    \left\langle v_{\textrm{rel}}\right\rangle = \sqrt{v_{0}^{2}-\frac{4\pi^{2}V^{2}_{0}}{\gamma^{2}_{b}d^{2}}}. \label{av-velocity-athermal_limit}
\end{equation}
The point $v_{0}^{\ast}=\frac{2\pi V_{0}}{\gamma_{b}d}$ marks the transition at which $\left\langle v_{\textrm{rel}}\right\rangle$ crosses over to a finite value. This transition is what we refer to as \textit{rupture transition} in the main text.

In Fig.~\ref{fig-flowcurve_athermal}, we have plotted the flow-curve computed from the average relative velocity for the case $v_{0}>v_{0}^{\ast}$ for different values of bath friction coefficient $\gamma_{b}$. As $\gamma_{b}$ gets larger, the flow-curve approaches its classical limit.

\subsection{Driving with time-dependent velocity}
As in the main text, we take the a sinusoidal form for the driving velocity, i.e. $v_{0}(t) = v_{00} + v_{01}\,\cos(2\pi\omega_{0}t)$. Eq. \eqref{PT_noise-free} thus reads as
\begin{eqnarray}
	\gamma_{b}\,\dot{z}(t) &=& -\gamma_{b}v_{00} -\frac{2\pi}{d} V_{0} \sin\left(\frac{2\pi}{d} z(t)\right)   \nonumber \\
	&& -\gamma_{b}v_{01}\,\cos(2\pi\omega_{0}t) . \label{PT_noise-free_t-dependent}
\end{eqnarray}
The above equation corresponds to the overdamped motion of particle in a periodic potential under an external time-dependent driving. These kind of models have been studied in both limits of damping, under- and over-damped, e.g. in the context of stochastic resonance~\cite{kim1998SR, saikia2011SR, reenbohn2015SR}. In the present work, we do not wish to make any attempt to study stochastic resonance, which itself is a subject of immense scientific research, and focus on understanding the observed oscillations in friction forces when the particle is dragged with a time-dependent velocity (see Fig~\ref{fig-force_vs_frequency-linear_response} ). 

Treating the time-dependent part in dragging velocity as perturbation, we first write the solution to Eq.~\eqref{PT_noise-free_t-dependent} up to linear order in $v_{01}$ as
\begin{equation}
    z(t) = z_{0}(t) + v_{01}\,z_{1}(t)\, , \label{solution-form}
\end{equation}
where $z_{0}(t)$ is the solution for the unperturbed case, i.e. when $v_{01}=0$. Using this solution into ~\eqref{PT_noise-free_t-dependent}, followed by an expansion in powers of $v_{01}$ and finally collecting the terms with same powers in $v_{01}$ leads to the following equations
\begin{eqnarray}
    \gamma_{b}\,\dot{z}_{0}(t) &=& -\gamma_{b}v_{00} -\frac{2\pi}{d} V_{0} \sin\left(\frac{2\pi}{d} z_{0}(t)\right), \label{eqn_noise-free_zero-order} \\
    \gamma_{b}\,\dot{z}_{1}(t) &=& -\gamma_{b}\,\cos(2\pi\omega_{0}t) -\frac{4\pi^{2}}{d^{2}} V_{0} \cos\left(\frac{2\pi}{d} z_{0}(t)\right) z_{1}(t) \label{eqn_noise-free_first-order} \nonumber \\
    &&
\end{eqnarray}
It may be noted immediately that $z_{0}(t)$ is given by Eq.~\eqref{solution_noise-free}. The solution for the perturbed part, with the initial condition $z_{1}(t=0)=0$, is given as
\begin{equation}
    z_{1}(t) = -\int_{0}^{t}dt'\,\cos \left(2\pi\omega_{0}t'\right) e^{-\int_{t'}^{t}ds\,f(s)}, 
\end{equation}
where 
\begin{equation}
    f(t) = \frac{1}{\gamma_{b}} \frac{4\pi^2}{d^2}V_{0}\cos\left(\frac{2\pi}{d} z_{0}(t)\right). 
\end{equation}
Differentiating w.r.t. to $t$ leads to the following expression for the (relative) velocity
\begin{eqnarray}
    \dot{z}(t) &=& \dot{z}_{0}(t) + v_{01}f(t) \int_{0}^{t}dt'\,\cos \left(2\pi\omega_{0}t'\right)\, e^{-\int_{t'}^{t}ds\,f(s)} \nonumber \\
    && - v_{01}\cos(2\pi\omega_{0}t).
\end{eqnarray}
As demonstrated earlier, the total drag-force acting on the system is proportional to the average relative velocity. An integration over the period of driving-velocity (period = $1/\omega_{0}$) thus leads to
\begin{eqnarray}
    \left\langle \Delta F\right\rangle_t &=& \left\langle F\right\rangle_t - \gamma(v_0)v_0 \nonumber \\
    &=& \gamma_{b}v_{01}\omega_{0}\int_{0}^{1/\omega_{0}}dt\, \cos(2\pi\omega_{0}t)\, e^{-\int_{t}^{1/\omega_0}ds\,f(s)}, \nonumber \\
    \label{eqn:change_in_force}
\end{eqnarray}
where $\left\langle \Delta F\right\rangle_t$ is the change in the mean friction force with respect to the unperturbed state. 

Near the critical velocity, i.e. in the linear response regime, one may use a series expansion in $v_{00}$ around $v_{0}^{\ast}=\frac{2\pi V_{0}}{\gamma_{b}d}$ to write
\begin{equation}
    f(t) \simeq \frac{2\pi v^{\ast}_{0}}{d} \frac{1 + \frac{2\pi v^{\ast}_{0}}{d}t} {1 + \frac{2\pi v^{\ast}_{0}}{d}t + \frac{1}2\frac{2\pi v^{\ast}_{0}}{d}t^2}.
\end{equation}
In writing the above equation, we have used the trigonometric identity, $\cos2A = \frac{1-\tan^{2}A}{1+\tan^{2}A}$ and Eq.~\eqref{solution_noise-free} with $\alpha^2 = v_{00}^{2} - v^{\ast 2}_{0}$ as solution for the case with constant driving. Using the above into Eq.~\eqref{eqn:change_in_force} leads to 
\begin{equation}
    \left\langle \Delta F\right\rangle_t = \frac{\gamma_b v_{01}v^{\ast 2}_{0}}{\omega_{0}^{2}d^2 + 2\pi v^{\ast}_{0} \omega_{0}d + \frac{1}{2}(2\pi v^{\ast}_{0})^2}.
\end{equation}
The above is a monotonically decreasing function of $\omega_0$ which has the largest value at zero, i.e. $\lim\limits_{\omega_{0}\rightarrow 0}\left|\Delta \bar{F}\right| = \frac{\gamma_b v_{01}}{2\pi^2}$ and smallest value of zero in the limit $\omega_0\rightarrow\infty$.

On the other hand, in Fig.~\ref{fig-force_vs_frequency-shear_thinning}, we have used mean velocity that is much larger than the corresponding critical velocity, i.e. $v_{00} > v^{\ast}_{0}$. In other words, we are quite far from the linear response regime where the velocity of bath particle is almost negligible w.r.t. to the velocity of tracer particle. Under such conditions, one may use the approximation 
\begin{equation}
    \alpha^2 = v_{00}^{2} - v^{\ast 2}_{0} \simeq v_{00}^{2}.
\end{equation}
This leads to
\begin{equation}
    f(t) \simeq  \frac{1}{\gamma_{b}} \frac{4\pi^2}{d^2}V_{0}\cos\left(\frac{2\pi v_{00}t}{d} \right)
\end{equation}
The expression from Eq.~\eqref{eqn:change_in_force} thus simplifies to give
\begin{eqnarray}
    \left\langle \Delta F\right\rangle_t 
    &=& 
    \gamma_{b}v_{01}\omega_{0}\,e^{-\frac{v^{\ast}_{0}}{v_{00}} \sin\left(\frac{2\pi v_{00}}{\omega_{0}d} \right)} \nonumber \\
    && \int_{0}^{1/\omega_{0}}dt\, \cos(2\pi\omega_{0}t)\, e^{\frac{v^{\ast}_{0}}{v_{00}} \sin\left(\frac{2\pi v_{00}t}{d}\right)}.
\end{eqnarray}
At the \textit{lower} harmonics, i.e. $\omega_0 = \frac{1}{n}\frac{v_{00}}{d}$, the integral in the above equation becomes zero, which matches quite well with the friction force plotted in Fig.~\ref{fig-force_vs_frequency-shear_thinning}. Since $\frac{v^{\ast}_{0}}{v_{00}}\ll 1$, one can further expand the exponential inside the integration to get
\begin{equation}
   \left\langle \Delta F\right\rangle_t
    = \frac{\gamma_{b}v_{01}v^{\ast}_{0}\omega_{0}d} {\pi\left( v^{2}_{00}-\omega^{2}_{0}d^2\right)}\,\sin^{2}\left(\frac{\pi v_{00}}{\omega_0 d}\right)\, e^{-\frac{v^{\ast}_{0}}{v_{00}} \sin\left(\frac{2\pi v_{00}}{\omega_{0}d} \right)}.
\end{equation}

%
\end{document}